\documentclass[useAMS,usenatbib]{mn2e}

\usepackage[british]{babel}
\usepackage{graphicx,amssymb,amsmath,epsfig,rotating,fleqn}
\usepackage{color}
\usepackage{float}
\usepackage{rotating}
\usepackage{tikz}
\usepackage{lscape}
\usepackage{caption}

\title
  [Bayesian analysis of cosmic rays]
  {A Bayesian analysis of the 69 highest energy cosmic rays detected by the Pierre Auger Observatory}
\author
  [A.\ Khanin \& D.\ J.\ Mortlock]
  {Alexander Khanin$^{1}$\thanks{E-mail: ak2008@imperial.ac.uk}
  and 
  Daniel J.\ Mortlock$^{1,2}$
\vspace{3mm}\\
$^1$Astrophysics Group, Imperial College London, Blackett Laboratory,
  Prince Consort Road, London SW7 2AZ, U.K. \\
$^2$Department of Mathematics, Imperial College London, London SW7 2AZ, U.K.
   }

\begin{document}

\date{Accepted 2016 ?????? ??. 
  Received 2016 ?????? ??; in original form 2016 ???????? ??}

\pagerange{\pageref{firstpage}--\pageref{lastpage}} \pubyear{2016}

\maketitle

\label{firstpage}

\begin{abstract}
The origins of ultra-high energy cosmic rays (UHECRs) remain an open question. Several attempts have been made to cross-correlate the arrival directions of the UHECRs with catalogs of potential sources, but no definite conclusion has been reached. We report a Bayesian analysis of the 69 events from the Pierre Auger Observatory (PAO), that aims to determine the fraction of the UHECRs that originate from known AGNs in the Veron-Cety \& Veron (VCV) catalog, as well as AGNs detected with the Swift Burst Alert Telescope (Swift-BAT), galaxies from the 2MASS Redshift Survey (2MRS), and an additional volume-limited sample of 17 nearby AGNs. The study makes use of a multi-level Bayesian model of UHECR injection, propagation and detection. We find that for reasonable ranges of prior parameters, the Bayes factors disfavour a purely isotropic model. 
For fiducial values of the model parameters, we report 68\% credible intervals for the fraction of source originating UHECRs of $0.09^{+0.05}_{-0.04}$, $0.25^{+0.09}_{-0.08}$, $0.24^{+0.12}_{-0.10}$, and $0.08^{+0.04}_{-0.03}$ for the VCV, Swift-BAT and 2MRS catalogs, and the sample of 17 AGNs, respectively.
\end{abstract}



\section{Introduction}
\label{section:intro}

Cosmic rays (CRs) are highly accelerated protons and atomic nuclei, some of which enter the Solar system and reach the Earth. They are the most energetic particles observed in nature, with energies in the range $10^{9}\,{\rm eV}$ to $10^{21}\,{\rm eV}$ (see e.g. \citealt{Kotera2011}, \ \citealt{Stanev2011} for reviews). 
\par
A number of open scientific issues remain with respect to CRs, in particular ultra-high energy cosmic rays (UHECRs) with arrival energies $E_{\rm{arr}} \gtrsim 10^{19}\,{\rm eV}$. The study of UHECRs is complicated by the fact that they experience an abrupt cutoff in their energy spectrum at $\sim$ $4 \times 10^{19}\, {\rm eV}$, so that only small samples are available. The largest currently available sample is the 69 events with $E_{\rm{arr}} \geq 5.5\times 10^{19}\, {\rm eV}$ recorded by the Pierre Auger Observatory (PAO) between 2004 January 1 and 2009 December 31 (\citealt{PAO2010}). 
\par
One open issue in the study of UHECRs is the question of their sources. A number of candidates, such as active galactic nuclei (AGNs) and pulsars have been proposed, but studies have not been conclusive (see e.g. \citealt{Kalmykov2013} for a review). The question of UHECR origins can be studied by attempting to associate the arrival directions with their sources. While UHECRs are charged particles and therefore experience magnetic deflection as they propagate, they are sufficiently energetic that the total deflection is expected to be $\sim 2$ to  $\sim 10$ deg (e.g. \citealt{Medina1998,Sigl2004,Dolag2005}), so that some information about their points of origin should be retained.\par
Association of UHECRs with catalogs of potential sources is made possible by the fact that UHECRs with energies of $E \ga 5\times 10^{19}\, {\rm eV}$ are expected to have come from a limited radius of $\sim 100 \,{\rm Mpc}$. This radius is sometimes called the Greisen-Zatsepin-Kuzmin (GZK) horizon, and arises due to the fact that UHECRs at those energies scatter off the cosmic microwave background (CMB) radiation in a process known as the GZK effect (\citealt{Greisen66}, \citealt{ZK66}).  The mean free path of the GZK effect at high energies is a few ${\rm Mpc}$ and the energy loss in each collision is $20$-$50\%.$ The resultant attenuation is very rapid, and is the cause of the cutoff in the UHECR energy spectrum observed by both HiRes (\citealt{Abbasi2008a}) and PAO (\citealt{Abraham2008}).
\par
A number of attempts have been made to find correlations between UHECR arrival directions and catalogs of possible sources. Cross-correlation studies have been conducted with galaxy catalogs, such as the Two Micron All-Sky Survey (2MASS) Redshift Survey (2MRS) (\citealt{PAO2009,Abbasi2010}), as well as specific types of objects such as active galactic nuclei (AGNs) (\citealt{PAO2007,PAO2008,George2008,Peer2009,Watson2011}) and BL Lacertae objects (BL LAcs) (\citealt{Tinyakov2001}). Overall, no clear consensus has been reached. Different studies have reported different degrees of correlation, depending on the statistical approach, the UHECR sample, and the population of source candidates that was used. The most significant correlation was reported by the Pierre Auger Collaboration, between arrival directions of UHECRs with energies $E \ge 5.7 \times 10^{19}\,{\rm eV}$ and the positions of nearby AGNs (\citealt{PAO2007}). The result was supported by Yakutsk data (\citealt{Ivanov2009}), but not by HiRes (\citealt{Abbasi2008}) or the Telescope Array (\citealt{AbuZayyad2012}). A more recent analysis of a larger PAO sample has shown a weaker correlation than before (\citealt{PAO2010}). \par
The lack of consensus on these issues is partly due to the difficulty of analyzing such small sample sizes. Given the small size of the UHECR data sets, it is important to utilize as much of the available information as possible. This can be achieved by adopting a Bayesian methodology, that involves models of the relevant physical processes. The first steps to such a comprehensive Bayesian work have been made in the recent work of \cite{Watson2011} and \cite{Soiaporn2012}. 

\cite{Watson2011} analysed the 27 events that were analysed in \cite{PAO2007}, and derived a posterior for the fraction that originated from AGNs in the Veron-Cetty \& Veron (VCV) catalog (\citealt{Veron2006}). To do so, they used a two-component parametric model characterized by a source rate $\Gamma$ and a background UHECR rate $R$. The model assumed that the UHECR arrival directions are points drawn from a Poisson intensity distribution on the celestial sphere. The intensity distribution was obtained with a computational UHECR model. \cite{Watson2011} report strong evidence of a UHECR signal from the VCV AGNs.  They find a low AGN fraction that is consistent with \cite{PAO2010}. For fiducial values of the model parameters, they report a 68\% credible interval for the AGN fraction of $F_{\rm{AGN}}=0.15^{+0.10}_{-0.07}$.

\cite{Soiaporn2012} developed a multi-level Bayesian framework to attempt to associate the 69 UHECRs that were recorded at the PAO in the period 2004-2009 with 17 nearby AGNs catalogued by \cite{Goulding2010} (hereafter G10). They report evidence for a small but nonzero fraction of the UHECRs to have originated at the AGNs from G10, of the order of a few percent to 20\%.

We extend the formalism of \cite{Watson2011} with both a greater data set and a refined UHECR model. Following \cite{PAO2010}, we extend the analysis to two further source catalogs: AGNs from the Swift Burst Alert Telescope (Swift-BAT) (\citealt{Baumgartner2010}) and galaxies from 2MRS (\citealt{Huchra2012}). We also extend the analysis to the 17 AGNs from the G10 catalog.
\par
After discussing the UHECR and source data sets in Section~\ref{sec:data}, we explain our UHECR model in Section~\ref{sec:CRmodel}, discuss the statistical formalism of our Bayesian model comparison in Section~\ref{sec:StatForm}, and the application of the formalism to mock data sets in Section~\ref{sec:Sim}. The results of applying the formalism to the PAO data are discussed in Section~\ref{sec:PAORes}. Some aspects of our computational approach are described in Appendix~\ref{sec:LikeApp}, and some subtleties of our model comparison are explored in Appendix~\ref{sec:BayesApp}. We use a Hubble constant of $H_0=70\, \rm{km/s/Mpc}$ where required to convert between redshifts and distances.
\section{Data}
\label{sec:data}
\subsection{UHECR sample}
\label{sec:Exposure}
The sample of UHECR events that was used in this analysis were the 69 highest energy events recorded at the PAO between January 2004 and November 2009, as documented in \cite{PAO2010}. These are the events with observed energies $E_{\rm{obs}}$ above the threshold $E_{\rm{obs}} \geq E_{\rm{thres}} = 5.7 \times 10^{19} \, \rm{eV} $.\par
The PAO is a CR observatory located in Argentina, at a longitude of 69.5$^\circ$ W and a latitude 35.2$^\circ$ S. PAO is a hybrid observatory, which means that it uses both surface detection (SD) and fluorescent telescope detection (FD) of UHECRs. The observatory has SD plastic scintillators of a total area of 3000 $\rm{km^2}$ and 4 FD telescopes.
\par
\begin{figure*}

\begin{center}$
\arraycolsep=4pt\def\arraystretch{0.01}
\begin{array}{cc}

\includegraphics[width=75mm]{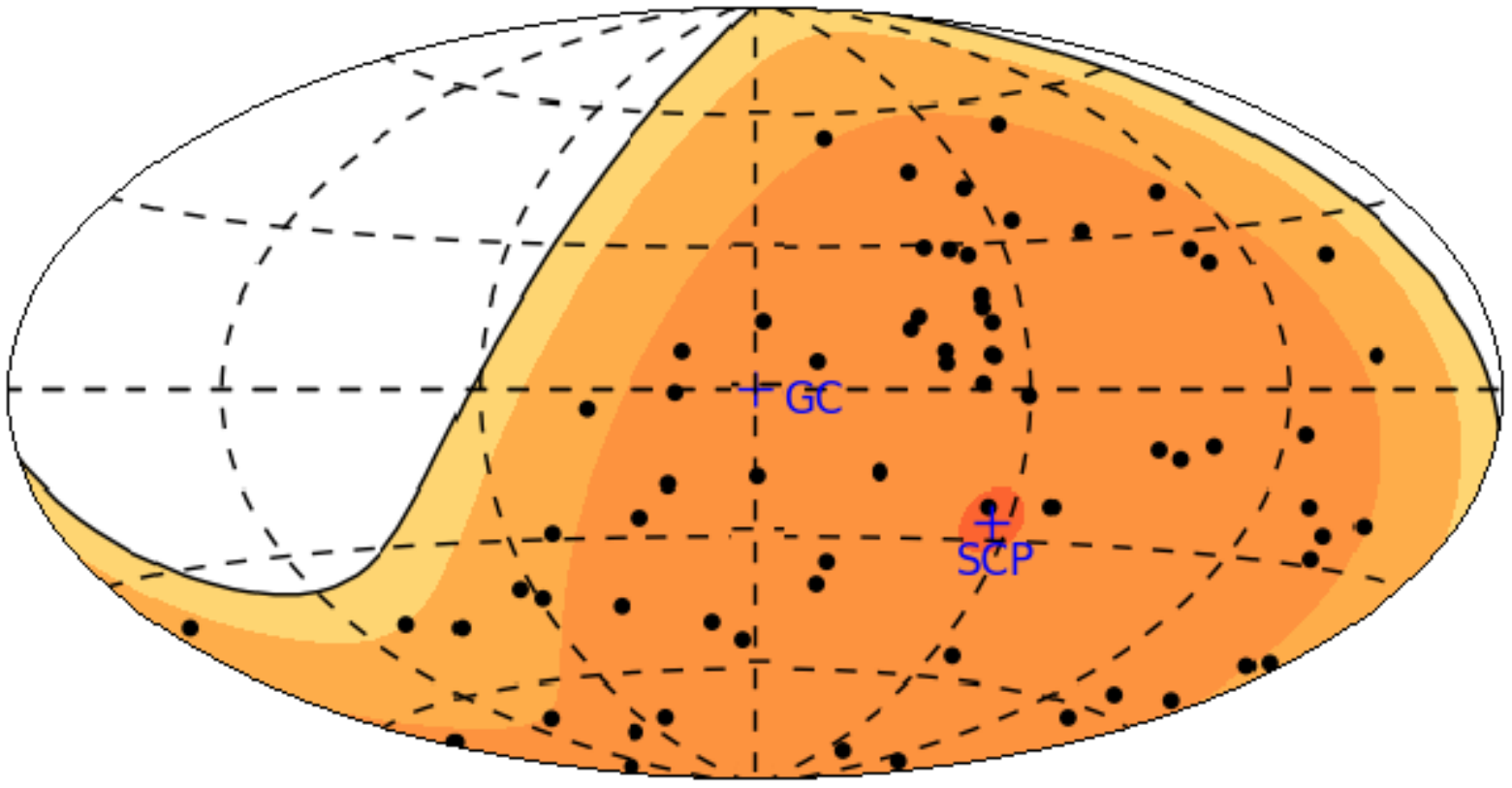}
\\
\includegraphics[width=75mm]{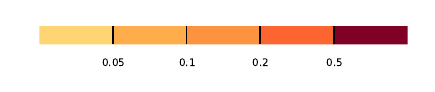}& \,
\end{array}$
\end{center}
\vspace{-5mm}
\caption{ Relative PAO exposure in Galactic coordinates. The arrival directions of the 69 UHECRs are shown as black points. The Galactic centre (GC) and south celestial pole (SCP) are indicated.}
\label{fig:TheExposure}
\end{figure*}

The PAO's total exposure of this data-set is $\epsilon_{{\rm tot}} = 20,370 \, \rm{km}^2 \, \rm{sr} \, \rm{yr} $ and its relative exposure per unit solid angle, ${\rm d}\epsilon / {\rm d}\Omega$, is illustrated in Figure~\ref{fig:TheExposure}.  The relative exposure is directly proportional to ${\rm  Pr}({\rm det}|\bmath{r})$, the probability that a UHECR will be detected if it arrives from direction $\bmath{r}$, but is normalized so that $\int ({\rm d}\epsilon / {\rm d}\Omega)\, {\rm d}\Omega = \epsilon_{{\rm tot}}$.  
\par
PAO measures UHECR arrival directions with an uncertainty of $\sim 1 \, \rm{deg}$ and arrival energies with a relative uncertainty of $\sim 12\%$ (\citealt{2013arXiv1310.4620L}).

\begin{figure*}
\begin{center}$
\arraycolsep=4pt\def\arraystretch{0.01}
\begin{array}{cc}
\text{(A) VCV } & \text{ (B) Swift-BAT} \\
\vspace{0.1cm} & \vspace{0.1cm} \\
\includegraphics[width=75mm]{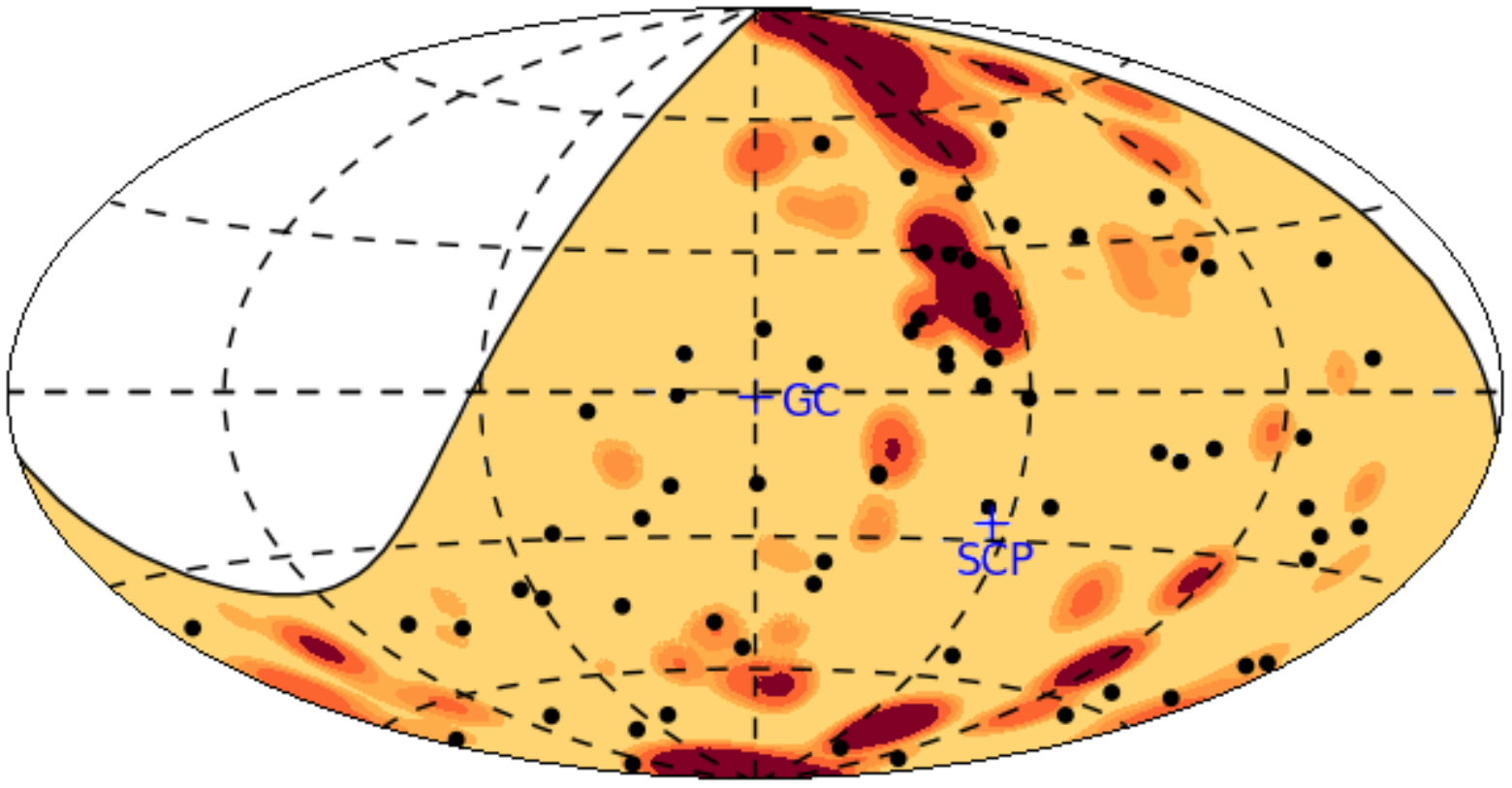}&\includegraphics[width=75mm]{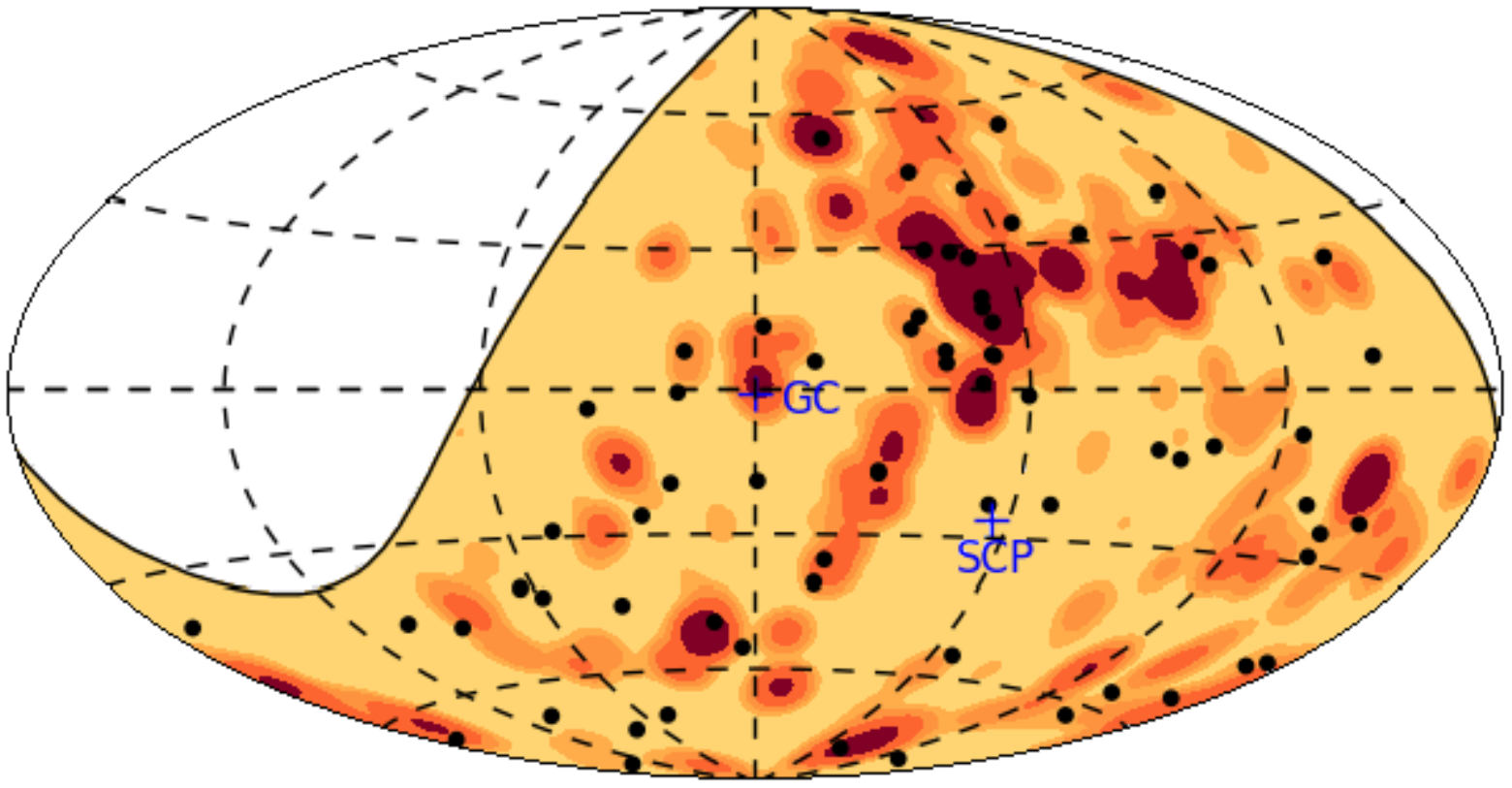}\\
\vspace{0.2cm} & \vspace{0.2cm} \\
\text{(C) 2MRS} & \text{(D) G10 } \\
\vspace{0.1cm} & \vspace{0.1cm} \\
\includegraphics[width=75mm]{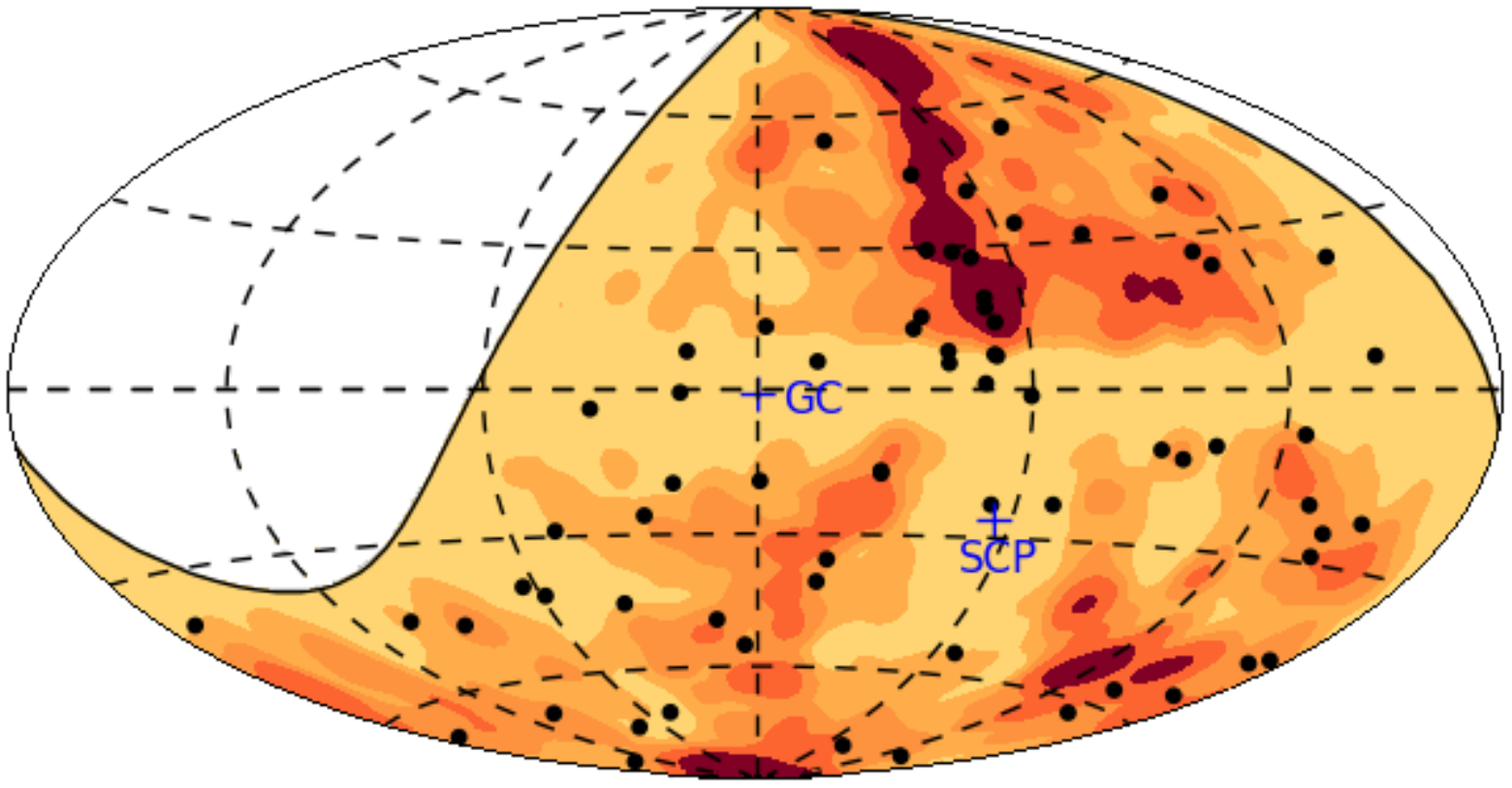}&\includegraphics[width=75mm]{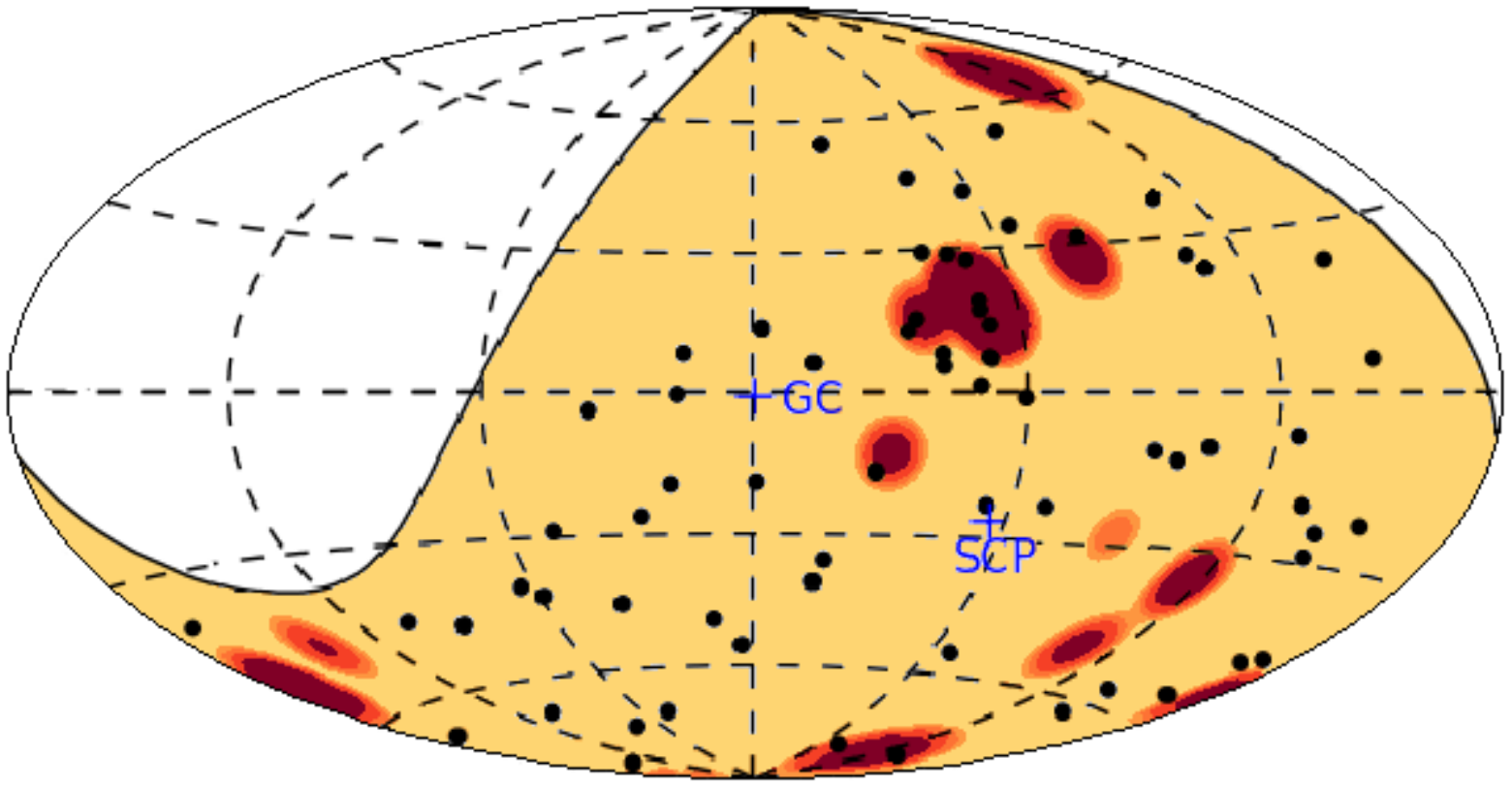}\\
\includegraphics[width=75mm]{Figures/Colorbar_6Oct.pdf}& \,
\end{array}$
\end{center}

\caption{ Positional dependence of the expected number of source originating events, for the VCV, Swift-BAT, 2MRS, and G10 catalogs. A fiducial value of the smearing parameter $\sigma=3 \, \rm{deg}$ is assumed. The arrival directions of the 69 UHECRs are shown as black points. Galactic coordinates are used, and the Galactic centre (GC) and south celestial pole (SCP) are indicated. }
\label{fig:TriModel}
\end{figure*}
\subsection{Source catalogs}
As potential source catalogs, we consider AGNs from the VCV, Swift-BAT and G10 catalogs, and galaxies from the 2MRS catalog. This allows us to compare our analysis for the Swift-BAT and 2MRS sources with the analysis from \cite{PAO2010}, our analysis for the VCV sources with the analyses from both \cite{PAO2010} and \cite{Watson2011}, and our analysis of the G10 sources with \cite{Soiaporn2012}.  
\par
We use the 12th edition of the VCV catalog, selecting sources with $z_{\rm{obs}}\leq 0.03$, as AGNs with higher redshift are too far away to be plausible UHECR sources, and can be shown to have a negligible effect on the results. We omit sources for which absolute magnitudes are not stated. The total number of VCV AGNs that meet those requirements is $N_{\rm{VCV}}=921$. This is the same sample of sources that was used in \cite{PAO2007},  \cite{PAO2010}  and \cite{Watson2011}, and in PAO's more recent analysis \cite{PAO2014}. While the VCV catalog is heterogenous and thus not ideal for statistical studies, it is close to complete for the low-redshift AGNs that are of relevance here.\par
\par
For the Swift-BAT catalog, we use the 58 month version, that includes a total of $N_{\rm{BAT}}=1092$ sources. In the case of the 2MRS catalog, we used the catalog version 2.4, 2011 Dec 16. We exclude events that are within $10^\circ$ of the Galactic plane, to avoid biases due to the incompleteness of the catalog in the region of the Galactic plane. This leaves a total of $N_{\rm{2MRS}}=20,702$ galaxies. These samples of Swift-BAT and 2MRS sources are the same as those used by \cite{PAO2010}. \par
The G10 catalog is a well-characterized volume-limited sample of AGNs. The 17 AGNs contained in it constitute all infrared-bright AGNs within 15 Mpc. This is the same sample that was used by \cite{Soiaporn2012}. 
\vspace{-1mm}
\section{UHECR model}
\label{sec:CRmodel}
A Bayesian UHECR analysis requires a realistic model of UHECR injection, propagation, and detection. This model was used both to compute the likelihoods in our statistical formalism (Section~\ref{sec:StatForm}), and to create simulated mock catalogs of UHECRs to test our methods (Section~\ref{sec:Sim}).
\subsection{Injection}
\label{sec:injection}
We adopt a model in which any given UHECR source emits UHECRs with an emission spectrum given by
\begin{equation}
\label{eq:erstmalrate}
\rm{d} \emph{N}_{emit} / \rm{d} \emph{E}_{emit}  \propto   \emph{E}_{\rm{emit}}^{-\gamma - 1} ,
\end{equation} 
where the logarithmic slope $\gamma$ is taken to be 3.6 (\citealt{Abraham2010}). The spectrum is normalized in such a way that the total emission rate of UHECRs with energy greater than $E_{\rm{emit}}$ is given by
\begin{equation}
\label{eq:erstmalrate}
\frac{\rm{d} \emph{N}_{\rm{emit}}(>\,\emph{E}_{\rm{emit}})}{\rm{d}\emph{t}}  =  \Gamma_s \left(\frac{E_{\rm{emit}}}{E_{\rm{min}}}\right)^{-\gamma} ,
\end{equation} 
where $E_{\rm{min}}=5.7 \times 10^{19} \, \rm{eV}$ is the minimum UHECR emission energy and $\Gamma_s$ is the rate at which source $s$ emits UHECRs with $E_{\rm{emit}} >E_{\rm{min}}$.

\subsection{Energy loss during propagation}
\label{sec:Eloss}

The energy loss processes experienced by UHECRs can be characterized in terms of the loss length        $L_{\rm{loss}} = -\emph{E}(\rm{d}\emph{E}/\rm{d}\emph{r})^{-1}$.
Given the loss length as a function of energy, it is possible to calculate the total amount of energy that a UHECR loses as it travels to the Earth from a given distance by solving the differential equation
\begin{equation}
\frac{\rm{d}\emph{E}}{\rm{d}\emph{r}} = -\frac{E}{L_{\rm{loss}}(E)}   .
\end{equation} 
For pure proton composition, $L_{\rm{loss}}$ obeys the expression
\begin{equation}
L_{\rm{loss}}^{-1}  =   \frac{1}{c}[\beta_{\rm{adi}}(E,z) +\beta_{\rm{GZK}}(E,z) + \beta_{\rm{BH}}(E,z) ]  ,
\end{equation} 
where $c$ is the speed of light and $\beta_{\rm{GZK}}(E,z)$, $\beta_{\rm{BH}}(E,z)$ and $\beta_{\rm{adi}}(E,z)$ are terms corresponding to the three main energy loss processes experienced by UHECRs of pure proton composition (e.g.\ \citealt{Stanev2009}):
\begin{enumerate}
\item
the GZK scattering off the CMB photons at energies above $E \ga 5\times 10^{19}\, {\rm eV}$;
\item
Bethe-Heitler (BH) ${\rm e^{+}e^{-}}$ pair production (also a scattering process off the CMB radiation), which dominates at lower energies (\citealt{Hillas1968});
\item
the adiabatic energy loss due to the expansion of the Universe.
\end{enumerate}
A detailed discussion of these terms, including expressions and parametrizations, can be found in \cite{Domenico2013}. For the energies that are relevant in this investigation, the dominant term is $\beta_{\rm{GZK}}(E,z)$. The Bethe-Heitler and adiabatic processes dominate the energy loss at lower energies, but play only a minor role at the higher energies in question.

\begin{figure}
$
\arraycolsep=0.01pt\def\arraystretch{-0.5}
\begin{array}{ccc}
\includegraphics[width=90mm]{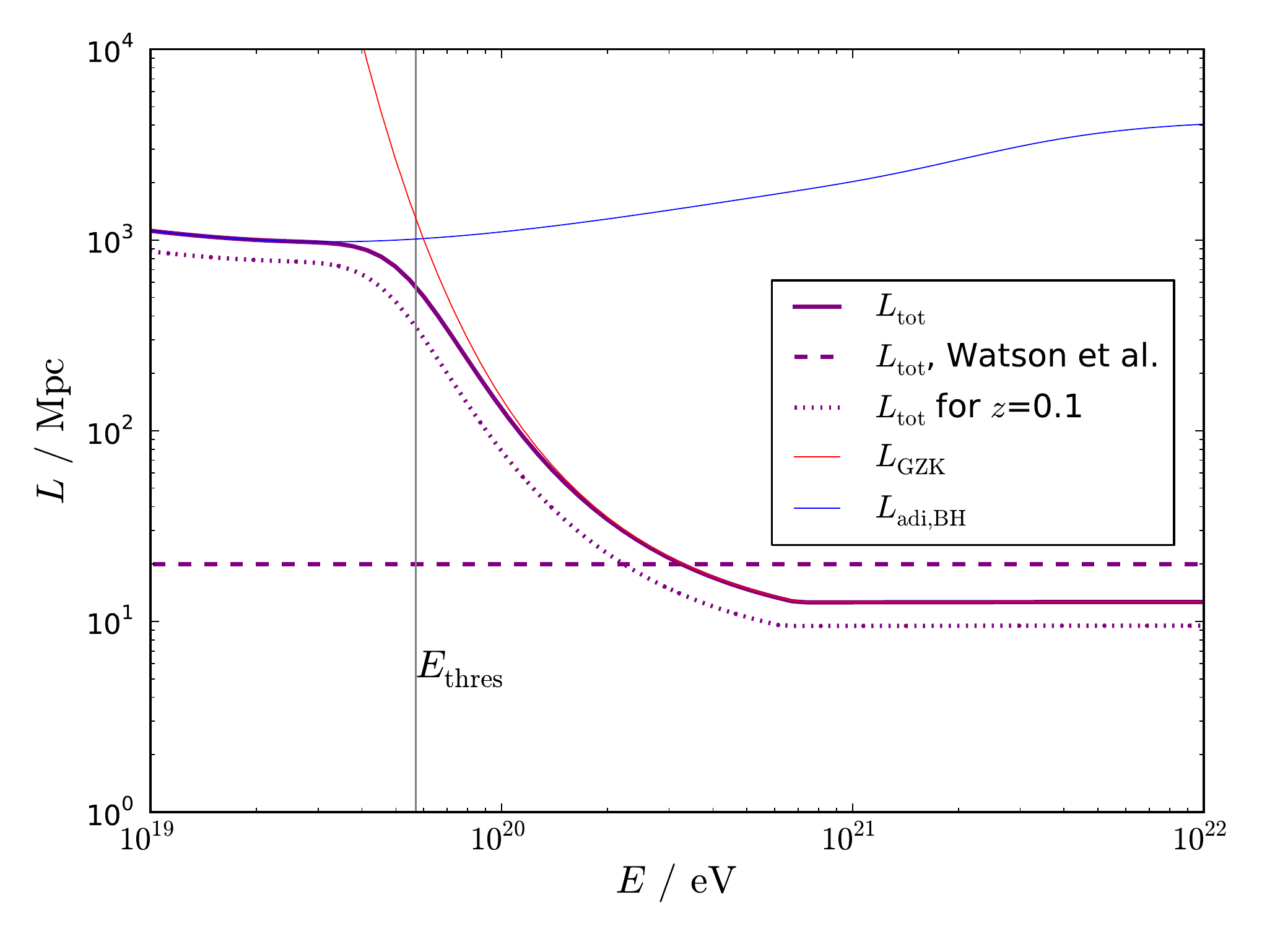}&
\end{array}$
\caption{ Loss lengths from the three energy loss processes, compared to the constant constant loss length used by \protect\cite{Watson2011}, as described in Section~\protect\ref{sec:Eloss}.  }
\label{fig:LossLengths}
\end{figure}

The loss lengths are shown as a function of energy in Figure~\ref{fig:LossLengths}. The contributions to the loss length from the BH and adiabatic losses are combined into a single function $L_{\rm{adi},\rm{BH}}$ that is contrasted with the loss length due to the GZK effect, $L_{\rm{GZK}}$. The two are combined into the total loss length $L_{\rm{tot}}$. The figure shows $L_{\rm{tot}}$ plots for $z$ values of 0.0 and 0.1, which correspond to distances of $0$ and $\sim400$ $\rm{Mpc}$, thus covering the GZK horizon. $L_{\rm{GZK}}$ appears very rapidly after an energy of $\sim 4\times 10^{19} \, \rm{eV}$ and begins to dominate the energy loss. As we are interested only in UHECRs with energies $E_{\rm{obs}} > E_{\rm{thres}}=5.7\times10^{19} \, \rm{eV}$, the GZK scattering is the most relevant loss process in this investigation.

The energy dependence of $L_{\rm{loss}}(E)$ is one of the main improvements of this propagation model over the model used in \cite{Watson2011}, where $L_{\rm{loss}}$ was taken to be a constant. The constant value of $L_{\rm{loss}}$ used by \cite{Watson2011} is also displayed in Figure~\ref{fig:LossLengths} for comparison. 

\subsection{Effective smearing}
\label{sec:smearing}

We combined the magnetic deflection that a UHECR experiences during propagation and the uncertainty in its detected arrival direction into a single kernel, which was chosen to be a von Mises-Fisher (vMF) distribution, defined as
\begin{equation}
    \label{eq:smtng}
{\rm Pr}(\bmath{\hat{r}} |\bmath{\hat{r}}_{\rm{src}}, \kappa)= \frac{\kappa}{4\rmn{\pi}\sinh(\kappa)} \rm{exp}(\kappa \bmath{\hat{r}} \cdot \bmath{\hat{r}}_{\rm{src}}),
\label{eq:Fish}
\end{equation}
where $\bmath{\hat{r}}$ is the measured arrival direction of the ray, $\bmath{\hat{r}}_{\rm{src}}$ is the source direction and $\kappa$ is the concentration parameter. The vMF distribution resembles a Gaussian on the sphere, with $\kappa$ being inversely related to the width of the Gaussian: for large values of $\kappa$ the distribution is peaked over an angular scale of $\sim 1 / \sqrt{\kappa}$ ;  if $\kappa$ tends to $0$ the distribution becomes uniform on the sphere.
\par
The magnitude of the deflection that the highest energy UHECRs experience is uncertain, with the estimates of typical deflection angles ranging from  $\sim 2$ to  $\sim 10$ deg (e.g. \citealt{Medina1998,Sigl2004,Dolag2005}). We assume a fiducial smearing angle of $\sigma \simeq$ 3 deg ($\kappa=360$), but also conduct investigations for smearing angles of $\sigma \simeq$ 6 and 10 deg ($\kappa=90$ and $30$).

\subsection{Observed UHECR flux}
\label{sec:ObsFlux}

The number of UHECRs from source $s$ above a threshold energy $E_{\rm{thres}}$ observed on Earth per unit area per unit time, $\rm{d} \emph{N}_{\emph{s}}(\emph{E}_{\rm{obs}} \geq \emph{E}_{\rm{thres}}) / \rm{d}\emph{t}\,\rm{d}\emph{A}$, is a quantity that is important in our statistical analysis. This rate is proportional to the rate of UHECRs emitted by the source, $\Gamma_{s}$, but it also depends on the distance-dependence of the UHECR energy loss, and on the UHECR injection spectrum. We use the UHECR propagation model described in Section~\ref{sec:Eloss} to determine the injection energy corresponding to the threshold energy $E_{\rm{thres}}$ and to the source distance $D_{s}$. Combining this value with Equation~\ref{eq:erstmalrate} and with the source distance $D_{s}$, we obtain
\begin{equation}
\label{eq:nochmalrate}
\frac{\rm{d}\emph{N}_{\emph{s}}(\emph{E}_{\rm{obs}}\,\geq \,\emph{E}_{\rm{thres}})}{\rm{d}\emph{t} \, \rm{d}\emph{A}}  =  \frac{\Gamma_s}{4\pi D_{s}^2} \bigg[\frac{E_{\rm{emit}}(E_{\rm{thres}})}{E_{\rm{min}}}\bigg]^{-\gamma}  .
\end{equation} 
This expression assumes that the observed energy $E_{\rm{obs}}$ is equivalent to the arrival energy of the UHECR, $E_{\rm{arr}}$. Thus, for the purposes of the calculation, the 12\% energy uncertainty of the PAO measurements is neglected. The variation in source rates $\Gamma_s$ among the sources that we are considering is not negligible. We use the source rate of Centaurus A as the reference value $\Gamma$. The source rate of a source $s$ is obtained by weighing the flux $F_s$ of that source in a particular band against the flux $F_{\rm{Cen}}$ of Centaurus A in that same band. The wave band of the flux thereby is different depending on the source catalog. For VCV, the flux of the source in the $V$-band is used, for Swift-BAT the X-ray flux, for 2MRS the IR flux, for G10 the $K$-band flux. The fluxes are thus used as weights, so that sources with higher flux contribute more UHECRs. This approach is very similar to the approach used in \cite{PAO2010}, where fluxes were used to weigh the sources from the Swift-BAT and 2MRS catalogs in the same way. Incorporating the fluxes into the formalism, we obtain the expression
\begin{equation}
\label{eq:nochmalrate}
\frac{\rm{d}\emph{N}_{\emph{s}}(\emph{E}_{\rm{obs}}\,\geq \,\emph{E}_{\rm{thres}})}{\rm{d}\emph{t} \, \rm{d}\emph{A}} = \frac{\Gamma}{4\pi D_{\rm{Cen}}^2}   \frac{ F_s }{ F_{\rm{Cen}} }    \bigg[\frac{E_{\rm{emit}}(E_{\rm{thres}})}{E_{\rm{min}}}{\bigg],}^{-\gamma}  
\end{equation} 
where $D_{\rm{Cen}}$ is the distance to Centaurus A.

\section{Statistical formalism}
\label{sec:StatForm}

Given a sample of UHECRs arrival directions, we would like to determine the fraction of these rays that have come from a set of sources under consideration. To do so, we use a two-component parametric model characterized by two rates:  The source rate $\Gamma$ and the isotropic background rate $R$. As elaborated in Section~\ref{sec:ObsFlux}, we use the source rate of Centaurus A as the reference value of $\Gamma$. We obtain a joint posterior distribution for the two rates:


\begin{equation}
\label{eq:posterior}
\rm{Pr(} \Gamma,\emph{R} | \textbf{\emph{d}}   \rm{)} = \mbox{} \frac{\rm{Pr(} \Gamma,\emph{R} \rm{)} \, \rm{Pr(} \textbf{\emph{d}} |\, \Gamma,\emph{R} \rm{)}}{\int_{0}^{\infty}\int_{0}^{\infty}\rm{Pr(} \Gamma,\emph{R} \rm{)} \, \rm{Pr(} \textbf{\emph{d}} |\, \Gamma,\emph{R}\rm{)} \, \rm{d}\Gamma \, d\emph{R} }    ,
\end{equation} 
where $\rm{Pr(} \Gamma,\emph{R} \rm{)}$ is the prior distribution for $\Gamma$ and $\emph{R}$, and $\rm{Pr(} \textbf{\emph{d}} |\, \Gamma,\emph{R} \rm{)}$ is the likelihood (i.e. the probability of obtaining the data set $\textbf{\emph{d}}$ given values of $\Gamma$ and $\emph{R}$). 

\subsection{Prior}

We adopt a uniform prior over $\Gamma$  and $\emph{R}$, with $\Gamma \! \geq \! 0$, $\emph{R} \! \geq \! 0$. This plausibly encodes our ignorance of the two parameters, and, unlike maximum entropy priors, includes a possible value of 0 for both parameters. The maximum values of $\Gamma$ and $R$ are denoted as $\rm{\Gamma_{max}}$ and $R_{\rm{max}}$. We have conducted our analysis for flat priors of varying width, using a variable width parameter $s$. The expression for the prior can be written as
\begin{equation}
\label{eq:priorprior}
{\rm Pr}(\Gamma, R|\textbf{\emph{d}}, M_2)  =  \frac{ 1 }{ s^2 \rm{\Gamma_{max}}  \emph{R}_{\rm{max}} } .
\end{equation}
$\rm{\Gamma_{max}}$ and $R_{\rm{max}}$ have been chosen in such a way that when $s=1$, the prior covers the 99.7\% credible region implied by the likelihood and an infinitely broad uniform prior. This gives a data driven scaling for the rates. The priors and their dependence on $s$ are illustrated in Appendix~\ref{sec:BayesApp}.  

\subsection{The likelihood}

To compute the likelihood, we use a `counts in cells' approach, in which the sky is divided into $1800 \times 3600 =$ 6,480,000 pixels, that are distributed  uniformly in right ascension and declination. Thus, the data set $\textbf{\emph{d}}$ can be rewritten as a set of counts in each pixel $\{{ \emph{N}_{\emph{\rm{c},\emph{p}}}}\}$.\par

The likelihood $\rm{Pr(}\textbf{\emph{d}} |\, \Gamma,\emph{R} \rm{)}$ is then given by a product of the individual Poisson likelihoods in each pixel, and can be written as
\[
\rm{Pr(} { \textbf{\emph{d}}  | \Gamma,\emph{R} }\rm{)}
\]
\begin{equation}
\label{eq:TotalLikelihood}
\mbox{} = \prod^{\emph{N}_{\rm{p}}}_{\emph{p}=1} \frac{(  \overline{\emph{N}}_{\rm{src},\emph{p}}+\overline{\emph{N}}_{\rm{bkg},\emph{p}})^{\emph{N}_{\rm{c},\emph{p}}} \exp[-(\overline{\emph{N}}_{\rm{src},\emph{p}}+ \overline{\emph{N}}_{\rm{bkg},\emph{p}} )]  }{\emph{N}_{\rm{c},\emph{p}}!}    ,
\end{equation} 
where $\overline{\emph{N}}_{\rm{src},\emph{p}}$ and $\overline{\emph{N}}_{\rm{bkg},\emph{p}}$ are the expected counts in pixel $\emph{p}$ due to sources and background, respectively. 
The expected number of counts in pixel $\emph{p}$ that are contributed by the background is 
\par
\begin{equation}
\label{eq:BackExp}
\overline{\emph{N}}_{\rm{bkg},\emph{p}}  =  \emph{R}\int_{p}\frac{\rm{d}\epsilon}{\rm{d}\Omega} \, \rm{d}\Omega_{\rm{obs}}  ,
\end{equation}
where the integral is over the pixel $\emph{p}$, and $\rm{d} \epsilon / \rm{d} \Omega $ is the relative exposure (Section~\ref{sec:Exposure}). 
The expected number of source originating events in pixel $\emph{p}$ is
\begin{equation}
\label{eq:SourceExp}
\overline{\emph{N}}_{\rm{src},\emph{p}} = \sum_{s=1}^{N_{\rm{s}}}\frac{\rm{d} \emph{N}_{\emph{s}}(\emph{E}_{\rm{obs}} \geq \emph{E}_{\rm{thres}})}{\rm{d}\emph{t}\,\rm{d}\emph{A}} \!\!\! \int_{p}\frac{\rm{d}\epsilon}{\rm{d}\Omega} \rm{Pr(} \vec{r}_{\rm{obs}} | \vec{r}_{\emph{s}} \rm{) \, \rm{d}\Omega_{\rm{obs}}}  ,
\end{equation} 
where the sum is over the sources, $\rm{Pr(} \vec{r}_{\rm{obs}} | \vec{r}_{\emph{s}} \rm{)}$ is the vMF distribution (Equation~\ref{eq:Fish}), and $\rm{d} \emph{N}_{\emph{s}}(\emph{E}_{\rm{obs}} \geq \emph{E}_{\rm{thres}}) / \rm{d}\emph{t}\,\rm{d}\emph{A} $ is the observed UHECR flux discussed in Section~\ref{sec:ObsFlux}. Inserting Equations~\ref{eq:BackExp} and~\ref{eq:SourceExp} into Equation~\ref{eq:TotalLikelihood}, we arrive at the full likelihood. 
\par
The positional dependence of $\overline{\emph{N}}_{\rm{bkg},\emph{p}}$ follows the relative exposure of PAO, as shown in Figure~\ref{fig:TheExposure}. The positional dependence of $\overline{\emph{N}}_{\rm{src},\emph{p}}$ depends both on the PAO exposure and on the distribution of sources in the given catalog. Figure~\ref{fig:TriModel} shows the dependence for the four catalogs that are used in this study. The dependence is dominated by the distribution of local AGNs, by far the strongest source being Centaurus A ($l=309.5^\circ$, $b=19.4^\circ$), which previously studies (e.g. \citealt{PAO2007}) have suggested as the dominant UHECR source.
\par
The expression for the likelihood can be rearranged to reduce the total number of computations, as described in Appendix~\ref{sec:LikeApp}.

\subsection{The source fraction}

The source fraction\footnote{The source fraction $F_{\rm{src}}$ is equivalent to the AGN fraction $F_{\rm{AGN}}$ used in \cite{Watson2011} but now generalized to allow for non-AGN progenitors.} is defined as the fraction of the UHECRs that are expected to have originated at the sources in whichever catalog is under consideration and is given by
\begin{equation}
\label{eq:BayesTheorem}
F_{\rm{src}}(\Gamma, R) = \frac{\sum_{\emph{p}=1}^{N_{\rm{p}}} \overline{\emph{N}}_{\rm{src},\emph{p}}  }{  \sum_{\emph{p}=1}^{N_{\rm{p}}} \overline{\emph{N}}_{\rm{src},\emph{p}} + \overline{\emph{N}}_{\rm{bkg}.\emph{p}} }  .
\end{equation} 
The posterior for $F_{\rm{src}}$ can be calculated from the posterior over the rates as
\[
\rm{Pr(}  \emph{F}_{\rm{src}} | \textbf{\emph{d}}   \rm{)}
\]
\begin{equation}
\mbox{} = \int\limits_{0}^{\Gamma_{\rm{max}}} \int\limits_{0}^{R_{\rm{max}}} 
 \rm{Pr(} \Gamma ,\emph{R} | \textbf{\emph{d}} \rm{)} \, \delta_{D}{ [\emph{F}_{\rm{src}}- \emph{F}_{\rm{src}}(\Gamma,\emph{R}) ] \,  \rm{d}\Gamma \, \rm{d}\emph{R} } .
\end{equation} 
$\rm{Pr(}  \emph{F}_{\rm{src}} | \textbf{\emph{d}}   \rm{)}$ is insensitive to $R_{\rm{max}}$ and $\Gamma_{\rm{max}}$ provided they are sufficiently large.

\subsection{Model comparison}
\label{section:BF}

We would like to compare model $M_1$ where all the UHECRs are drawn from a uniform distribution with model $M_2$ where the UHECRs are derived from a  combination of a background and a source originating component. To do this, we conduct a Bayesian model comparison. For a data set $\textbf{\emph{d}}$, and two models $M_1$ and $M_2$, the ratio of the marginal likelihoods for the two models, termed the Bayes factor, is 
\begin{equation}
 \label{eq:BayesFactor0}
  B_{12} = \frac{{\rm Pr}(\textbf{\emph{d}}|M_{\rm 1})}{{\rm Pr}(\textbf{\emph{d}}|M_{\rm 2})}.
\end{equation}
\par
In the specific case that is considered here, the models are nested: When $\Gamma=0$, model $M_2$ reduces to model $M_1$. A general expression of the Bayes factor in this situation is
\begin{equation}
 \label{eq:BayesFactor}
  B_{12} = \frac { \int {\rm Pr}(R|M_1) \, {\rm Pr}(\textbf{\emph{d}} | R,M_1) \, \rm{d} \emph{R} }{ \int {\rm Pr}(\Gamma, R |M_2) \, {\rm Pr}(\textbf{\emph{d}} | \Gamma ,R ,M_2) \, \rm{d}\Gamma  \, \rm{d}\emph{R} }.
\end{equation}
It can be shown (Dickey 1971) that in the case of such nested models, the expression reduces to 
\begin{equation}
\label{eq:nochmalrate}
B_{12} = \frac {  {\rm Pr}(\Gamma = 0 |\textbf{\emph{d}}, M_2) }{  {\rm Pr}(\Gamma = 0| M_2) }. 
\label{eq:SDDR}
\end{equation}
This expression is known as the Savage-Dickey Density Ratio, or SDDR.
Qualitatively, this expression means that the nested uniform model is preferred if, within the context of the more complicated model, the data result in an increased probability that $\Gamma=0$.

\section{Simulations}
\label{sec:Sim}

In order to investigate the constraining power of a data set of 69 events, we apply the method to simulated data sets. We use two extreme cases:
\begin{enumerate}
\item
Uniform arrival directions. These rays were drawn from a probability distribution that followed the PAO exposure.
\item 
UHECRs originating at sources from a catalog. We conducted simulations for all four of the catalogs. In each catalog, the sources were weighted by their fluxes and the PAO exposure. Random sources were then selected, and the propagation model of Section~\ref{sec:CRmodel} was used to propagate rays from the sources to the Earth.
\end{enumerate}
\par
\begin{figure}
$
\arraycolsep=0.01pt\def\arraystretch{-0.5}
\begin{array}{ccc}
\includegraphics[width=90mm]{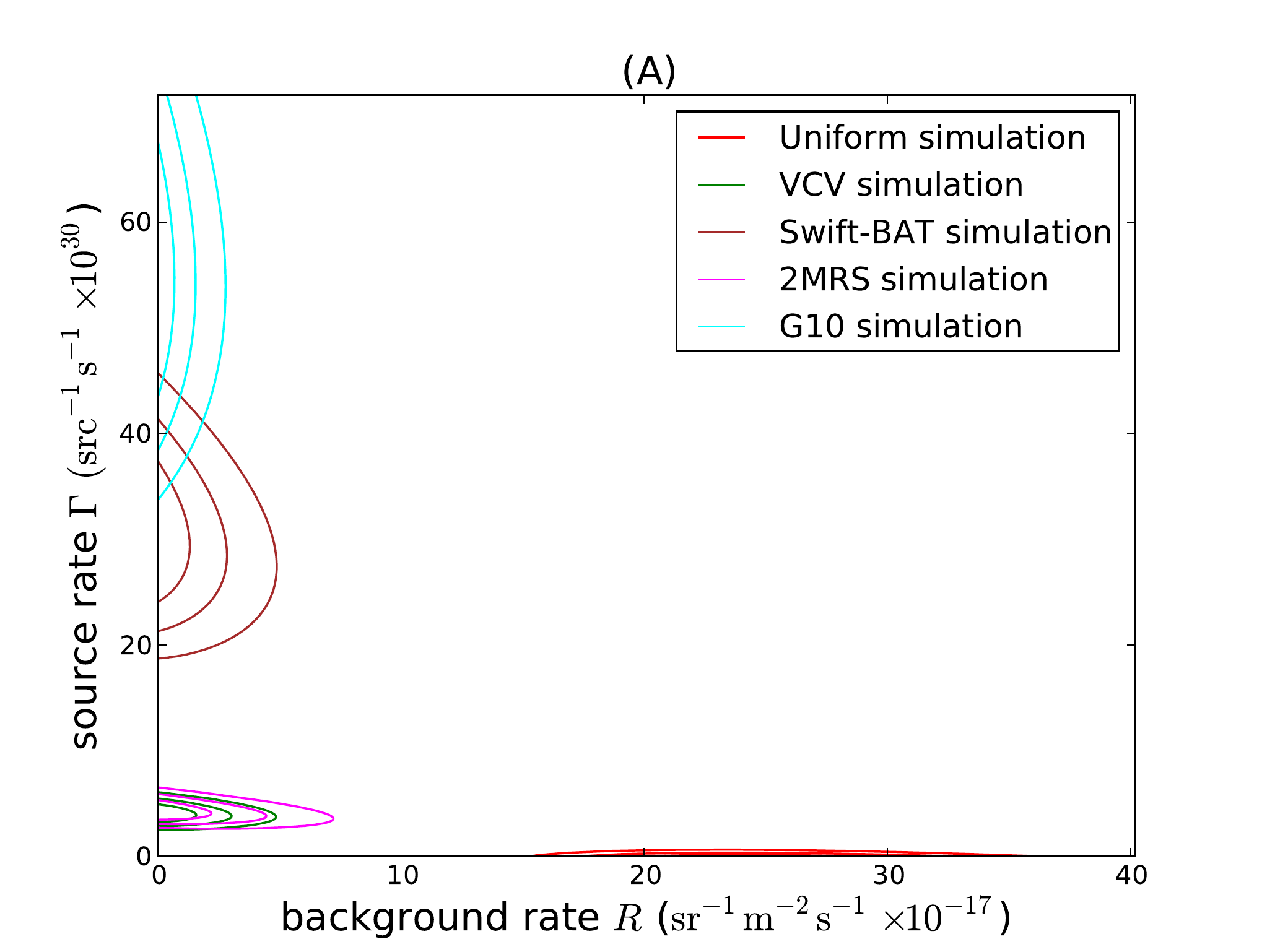} \\
\includegraphics[width=90mm]{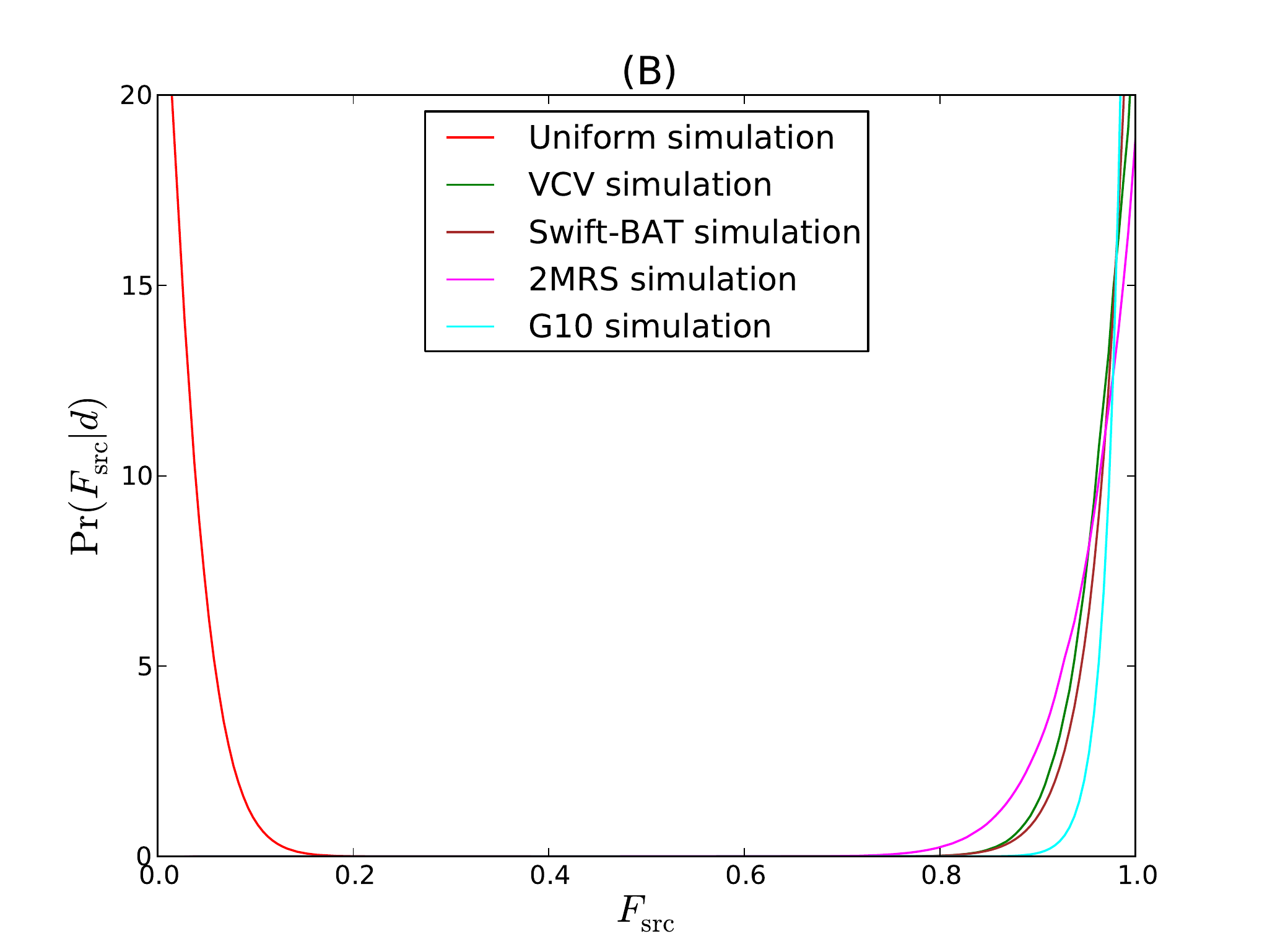} \\
\includegraphics[width=90mm]{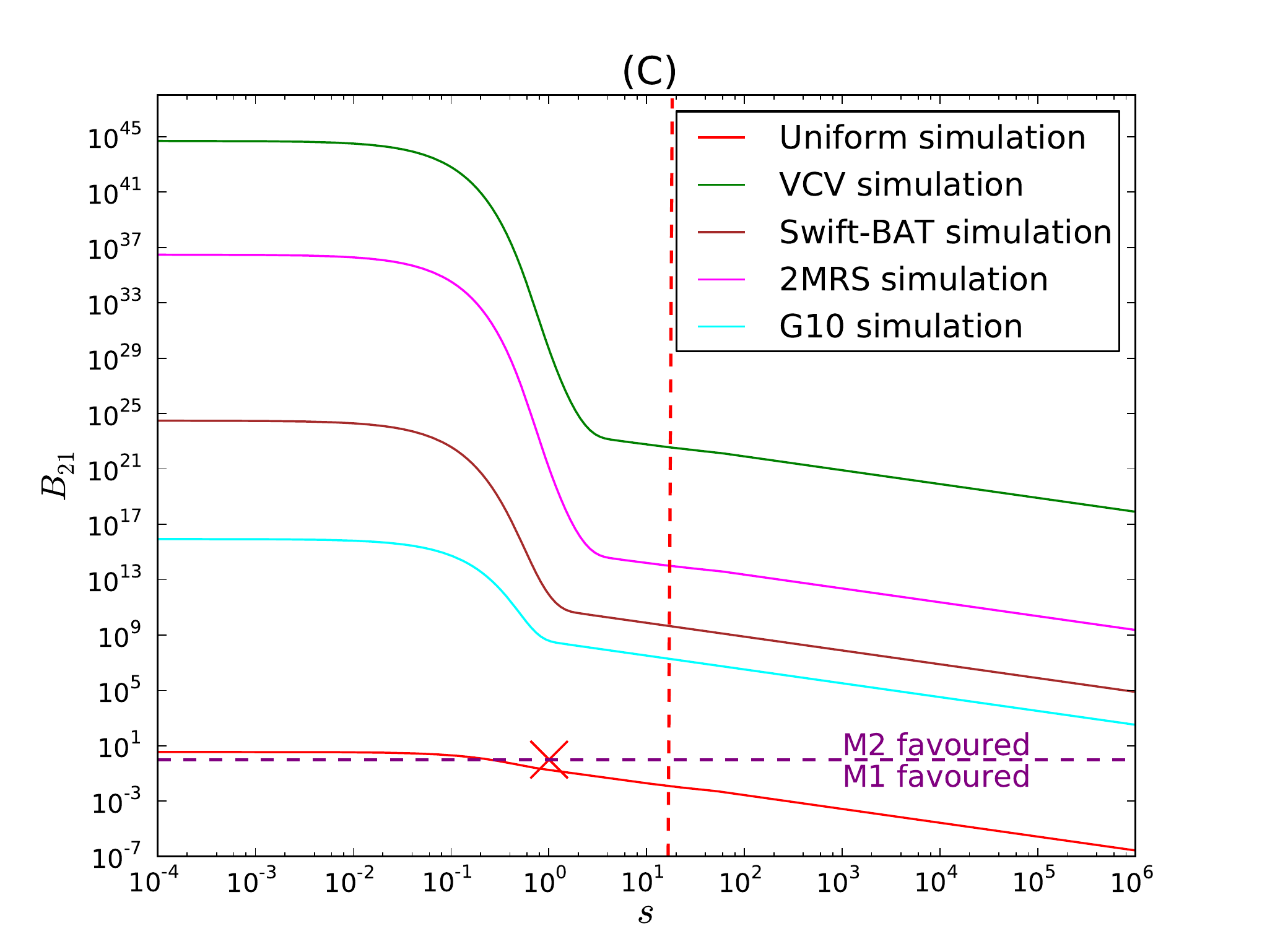}
\end{array}$
  \caption{Results from simulations: Uniform UHECRs, and UHECRs originating at sources from the 4 catalogs. In all cases, 69 events are used. (A) Posteriors for $\Gamma$ and $R$. The contours are the 68.3\%, 95.4\% and 99.7\% highest posterior density credible regions. (B) Posteriors for the source fraction. (C) Plot of Bayes factors $B_{21}$ as a function of the hyperparameter $s$. In (C), the $\times$-mark and the vertical line signify the minimum and the maximum values of the physically plausible range of $s$. The minimum and maximum values that are displayed correspond to the uniform simulation. }
\label{fig:UniAgn}
\end{figure}

The posteriors for the source and background rates, as well as the posteriors for the source fraction, are summarized in Figure~\ref{fig:UniAgn}. The posteriors for the uniform and source centred cases are completely disjoint, which demonstrates that in extreme senarios where all UHECRs originate either from a uniform background or from a source catalog, a data set of 69 events should be sufficient to distinguish between the two models. Figure~\ref{fig:UniAgn} also shows the Bayes factors as functions of $s$ for the two cases. The Bayes factors $B_{21}$ that are displayed are the inverses of the SDDR given in Equation~\ref{eq:SDDR}, and favour the more complex model for Bayes factors $>1$. \par
To assess the results of the Bayes factor simulations, we can derive a rough range of plausible values of $s$ from physical models, and then look at the behaviour of the Bayes factors at those physically plausible values. Plausible models of UHECR injection predict that the UHECR luminosity of a source like Centaurus A is of the order of $2.9 \times 10^{39} \, \rm{erg} \, \rm{s}^{-1}  \simeq  1.81 \times 10^{51} \, \rm{eV}$ (\citealt{Fraija2012}). If this is taken as the typical UHECR luminosity of a source, then for a UHECR energy range of $(5.7 - 100) \times 10^{19} \, \rm{eV}$, the range of source rates can be calculated by dividing the UHECR luminosity by the limiting values of this range. The result of this calculation is a range of source rates $\Gamma$ of roughly $(2 -  33) \times 10^{30} \, \rm{s^{-1}}$. The values of $s$ corresponding to this range have been marked on Figure~\ref{fig:UniAgn}. (The values are slightly different for each of the simulations. For the sake of clarity, only the values for the uniform simulation are displayed, the others being broadly similar.) For the sourced case, model $M_2$ is strongly favoured for all physically plausible values of $s$, while for the uniform case, the simple uniform model $M_1$ is favoured for the physically plausible values.

\section{Results}
\label{sec:PAORes}

The results of the application of the statistical methods described in Section~\ref{sec:StatForm} to the data described in Section~\ref{sec:data} are shown in Figures~\ref{fig:Blues} and~\ref{fig:ThreeCats}. Figure~\ref{fig:Blues} contrasts the results from our analysis with the equivalent results from \cite{Watson2011}, and with the results for an intermediate case. The use of a more refined propagation model leads to a higher posterior probability for lower source rates. The reason for that is that in Watson's propagation model, the energy loss length is constant and very small (Figure~\ref{fig:LossLengths}). UHECRs experience more drastic energy loss than in the more realistic model, which leads to more distant AGNs being excluded as plausible source candidates. As fewer sources are included, a higher source rate is required to generate the same sample of UHECRs. 
\par
The inclusion of 69 events reduces the extent to which the non-uniform model is favoured. This is evident from the posterior of the source fraction, and also from the behavaviour of $B_{21}$. This result agrees with the results of \cite{PAO2010}, which reported that the full 69 events yield lower evidence of anisotropy than the earlier study \cite{PAO2007}, which analysed 27 events.
\par
Figures~\ref{fig:ThreeCats} shows results for all four of the source catalogs, and for all values of the smearing parameter. Displayed are the posteriors for the source fraction, as well as plots of $B_{21}$ against $s$. The constraints on the source fraction for all cases are shown in Table~\ref{table:ConfInt}. The figures and table show that for greater smearing, the range of plausible values of $F_{\rm{src}}$ is increased, and the most probable value of the source fraction is higher than for the fiducial model of $\sigma=3 \, \rm{deg}$. The reason is that for greater magnetic deflection, the UHECR intensity distribution becomes more uniform, so that the uniform and mixed models become more difficult to distinguish, and a greater range of $F_{\rm{src}}$ values become viable. 
\par
The plots of $B_{21}$ demonstrate that for all physically plausible prior ranges of the model parameters, the fully isotropic model is disfavoured. The form of the dependence of $B_{21}$ on $s$ is elaborated upon in Appendix~\ref{sec:BayesApp}.
\par
These results for the VCV, Swift-BAT, and 2MRS catalogs can be compared with the results of \cite{PAO2010}, who used a correlation-based analysis on the VCV catalog that mirrored the analysis in \cite{PAO2007}. \cite{PAO2010} reported a correlation of (38$^{+7}_{-6}$)\% between UHECRs and sources from the VCV catalog, which was considerably lower than than the (69$^{+11}_{-13}$)\% correlation that was reported in \cite{PAO2007}. This reduction in the correlation is consistent with our findings that the source fraction is reduced as we increase the data set from 27 to 69 events. In addition to  these correlation based methods, \cite{PAO2010} conducted a likelihood based study similar to the analysis presented here, where the likelihood was taken as a probability map of arrival directions of UHECRs, parametrized by a magnetic smoothing angle $\sigma$ and a fraction of isotropic rays $f_{\rm{iso}}$, which is equivalent to $1-F_{\rm{src}}$. These likelihood-based studies were conducted for the Swift-BAT and 2MRS catalogs. For the 2MRS case, the maximum likelihood values of $f_{\rm{iso}}$ and $\sigma$ are reported as 0.56 and 7.8$^\circ$, respectively. The $\sigma$ value lies between our chosen smearing angles 6$ \, \rm{deg}$ and 10$ \, \rm{deg}$. The value for $f_{\rm{iso}}$ corresponds to a value of $F_{\rm{src}}$ of 0.44, which is consistent with our $F_{\rm{src}}$  credible intervals for these chosen smearing angles. For the case of Swift-BAT, the maximum likelihood value of $f_{\rm{iso}}$ is given as 0.64, which corresponds to a source fraction of 0.36. The maximum likelihood estimate of the smearing angle is reported as 1.5$^\circ$, which is lower than our minimum chosen value of 3$ \, \rm{deg}$. Despite the difference between the angles, a $F_{\rm{src}}$ value of 0.36 can still be considered broadly consistent with the 68\% credible interval for 3$ \, \rm{deg}$, $0.25^{+0.09}_{-0.08}$.
\par
Our results for the G10 catalog can be compared with the work of \cite{Soiaporn2012}. That analysis involved the full data set of 69 events, and found evidence for small but nonzero values of $F_{\rm{src}}$, of the order of a few percent to 20\%, ruling out values of $F_{\rm{src}} > 0.3$. This is broadly consistent with our results, which suggest that values of $F_{\rm{src}} <\sim 0.3$ are the most probable for all values of the smearing parameter.
{\renewcommand{\arraystretch}{1.5}
\begin{table}
\caption{Maximum a posteriori estimates and 68\% credible intervals for $F_{\rm{src}}$.} 
\begin{tabular}{c c c c} 
\hline\hline 
Catalog & $\sigma=3 \, \rm{deg}$ &$\sigma=6 \, \rm{deg}$ & $\sigma=10 \, \rm{deg}$ \\ [0.5ex] 
\hline 
VCV & $0.09^{+0.05}_{-0.04}$ & $0.14^{+0.07}_{-0.06}$ & $0.22^{+0.09}_{-0.08}$ \\ 
Swift-BAT & $0.25^{+0.09}_{-0.08}$ & $0.37^{+0.11}_{-0.10}$ & $0.46^{+0.13}_{-0.12}$ \\
2MRS & $0.24^{+0.12}_{-0.10}$ & $0.33^{+0.14}_{-0.14}$ & $0.40^{+0.15}_{-0.15}$ \\ 
G10 & $0.08^{+0.04}_{-0.03}$ & $0.14^{+0.06}_{-0.05}$ & $0.22^{+0.07}_{-0.07}$ \\[1ex] 
\hline 
\end{tabular}
\label{table:ConfInt} 
\end{table}
{\renewcommand{\arraystretch}{1.0}

\begin{figure}
$
\arraycolsep=0.01pt\def\arraystretch{0.01}
\begin{array}{c}
\includegraphics[width=90mm]{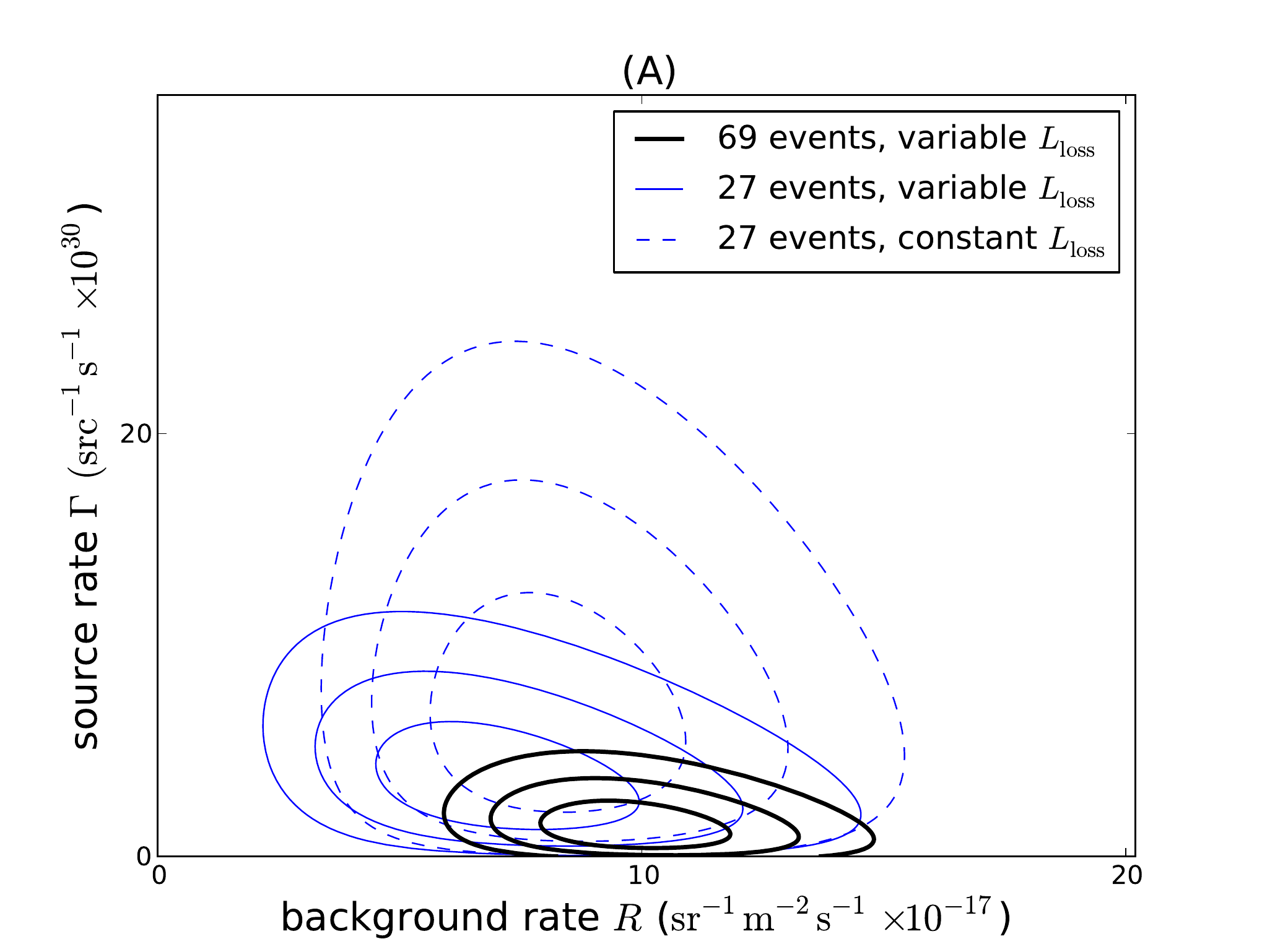}\\
\includegraphics[width=90mm]{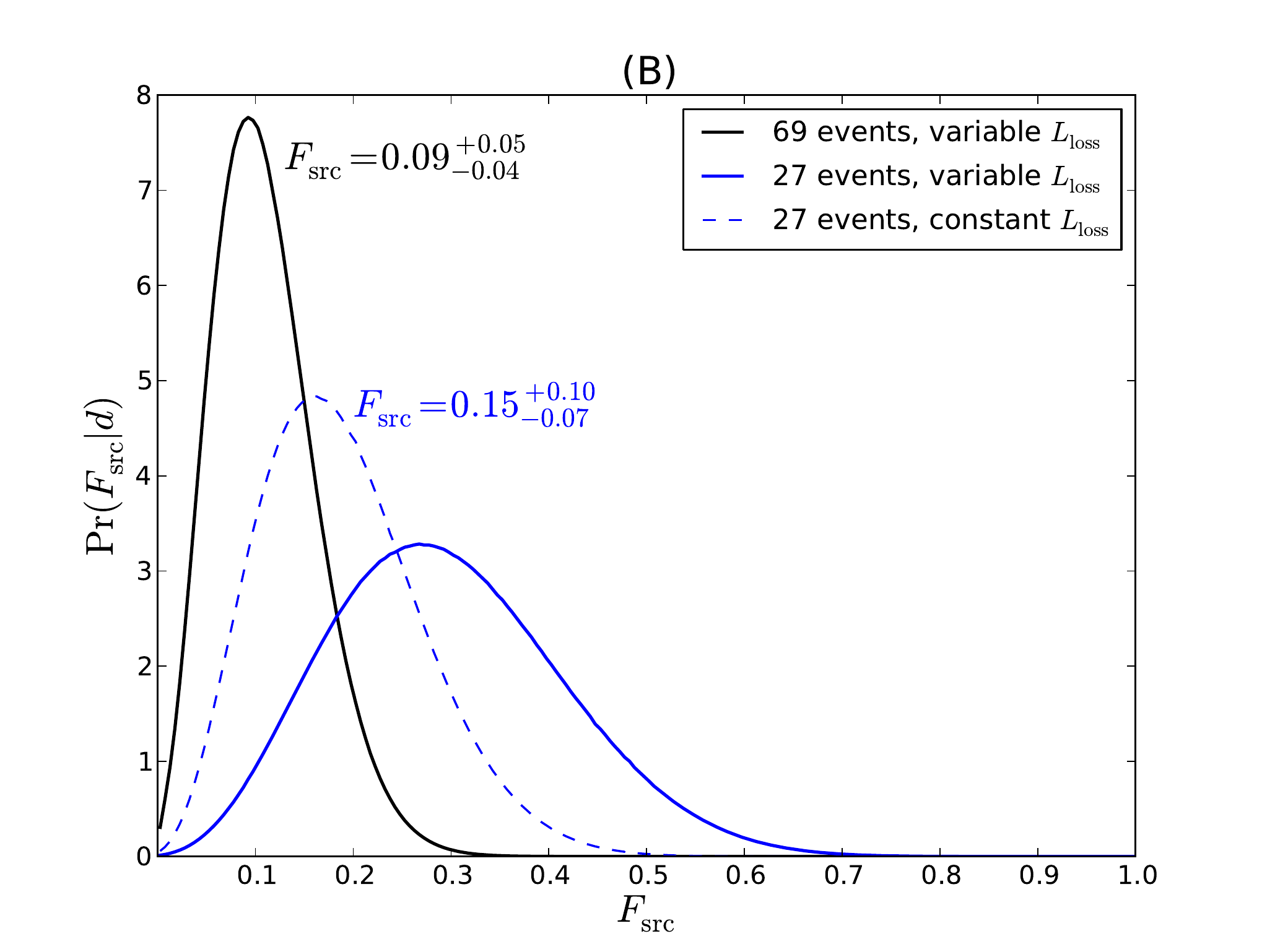} \\
\includegraphics[width=90mm]{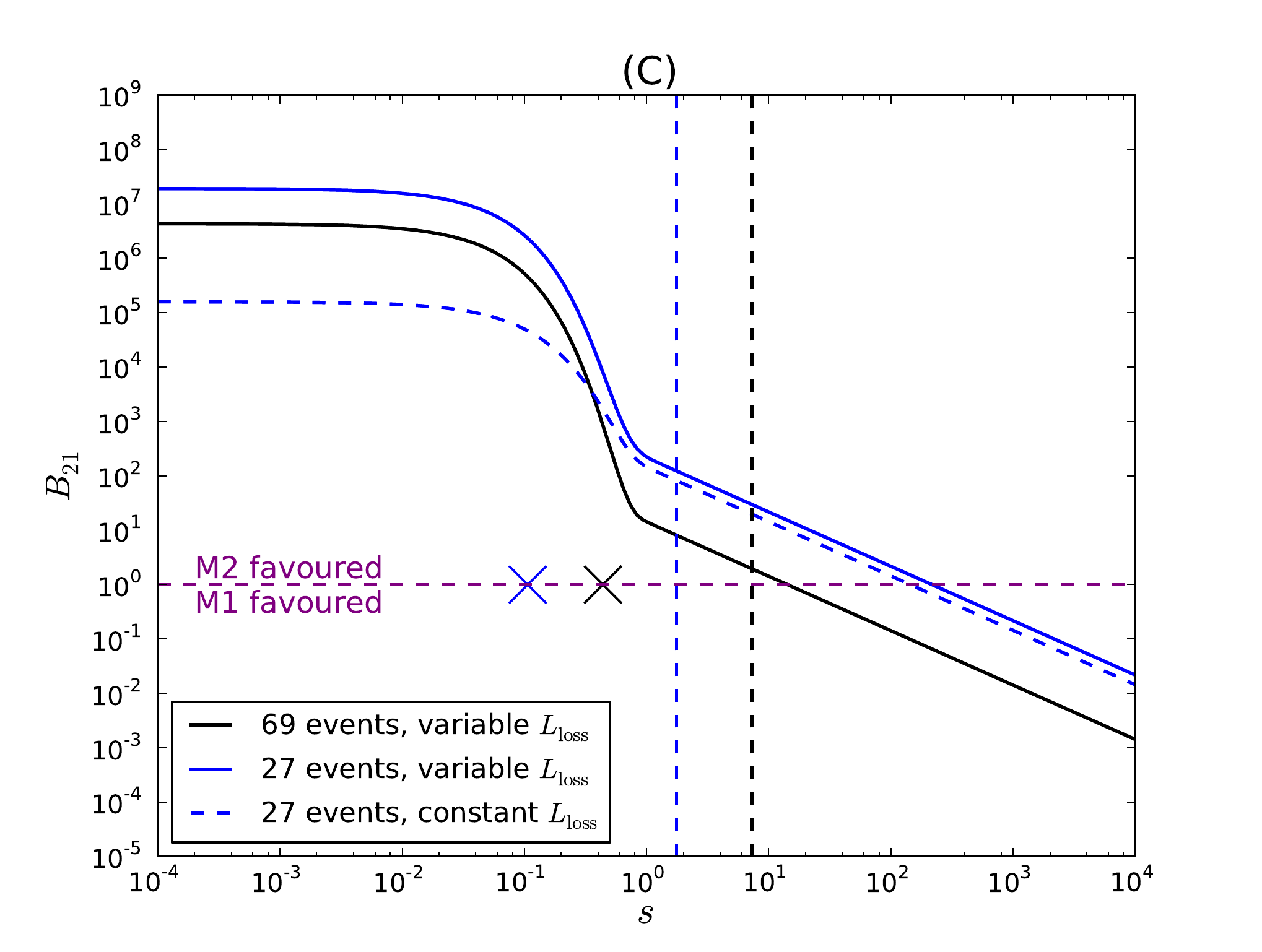}
\end{array}$
\caption{ Results for $\sigma=3 \, \rm{deg}$, and the sources from the VCV catalog. Results for 27 and 69 events, and for constant and variable loss lengths are displayed. (A) Posteriors for the source and background rates. The contours are the 68.3\%, 95.4\% and 99.7\% highest posterior density credible regions. (B) Posterior for the source fraction. (C) Plot of Bayes factors $B_{21}$ as a function of the hyperparameter $s$. In (C), physically plausible ranges of $s$ are shown for the cases of 27 events (blue) and 69 events (black), with a variable loss length. The $\times$-marks and the vertical lines signify the minimum and the maximum values of the physically plausible ranges of $s$. }
\label{fig:Blues}
\end{figure}

\section{Conclusions}
\label{sec:Conc}

We have performed a Bayesian analysis of the 69 UHECRs detected by the PAO with energies $E_{\rm{obs}}> 5.7 \times 10^{19}\,{\rm eV}$ to determine the fraction of these UHECRs that originated from catalogs of plausible UHECR sources. The sources considered were AGNs from the VCV, Swift-BAT, and G10 catalogs, and galaxies from the 2MRS catalog.
\par
For the fiducial magnetic smearing parameter of $\sigma =$ 3 deg, we report 68\% credible intervals for the source fraction of $0.09^{+0.05}_{-0.04}$, $0.25^{+0.09}_{-0.08}$, $0.08^{+0.04}_{-0.03}$ and $0.24^{+0.12}_{-0.10}$ for the VCV, Swift-BAT, G10 and 2MRS catalogs, respectively. For all physically plausible values of the model parameters, the fully uniform model is disfavoured. The results of our study are in broad agreement with previous work on this subject, such as \cite{Watson2011}, \cite{PAO2010} and \cite{Soiaporn2012}. The credible intervals for the VCV catalog are lower than the analogous credible intervals from \cite{Watson2011}, which used a similar method to analyse 27 PAO events. This is consistent with earlier studies: \cite{PAO2010}, which analysed 69 events, reported a lower signal of anisotropy than the earlier study \cite{PAO2007}, which used 27 events.
\par
We will extend this Bayesian framework to include the arrival energies of the UHECRs as well as the arrival directions.

It is expected that future experiments will produce data sets that will be sufficiently large for our Bayesian method (and other statistical approaches; see e.g.\ \citealt{FutureAnis2014}) to detect even the weak clustering expected if the UHECRS have come from nearby sources. PAO is continuing to take data and is expected to produce a sample of $\sim 250$ UHECRs over its first decade of operations.  Looking further ahead, the planned Japanese Experiment Module Extreme Universe Space Observatory (JEM-EUSO, \citealt{JEMEUSO2013}) on the International Space Station (ISS) is scheduled for launch in 2017 and is expected to detect $\sim 200$ UHECRs annually over its five year lifetime.

\par

\vspace{-0.5cm}

\bibliographystyle{mn2e}
\bibliography{bibli}

\begin{figure*}
\begin{center}$
\arraycolsep=0.01pt\def\arraystretch{0.01}
\begin{array}{cc}
\includegraphics[width=85mm]{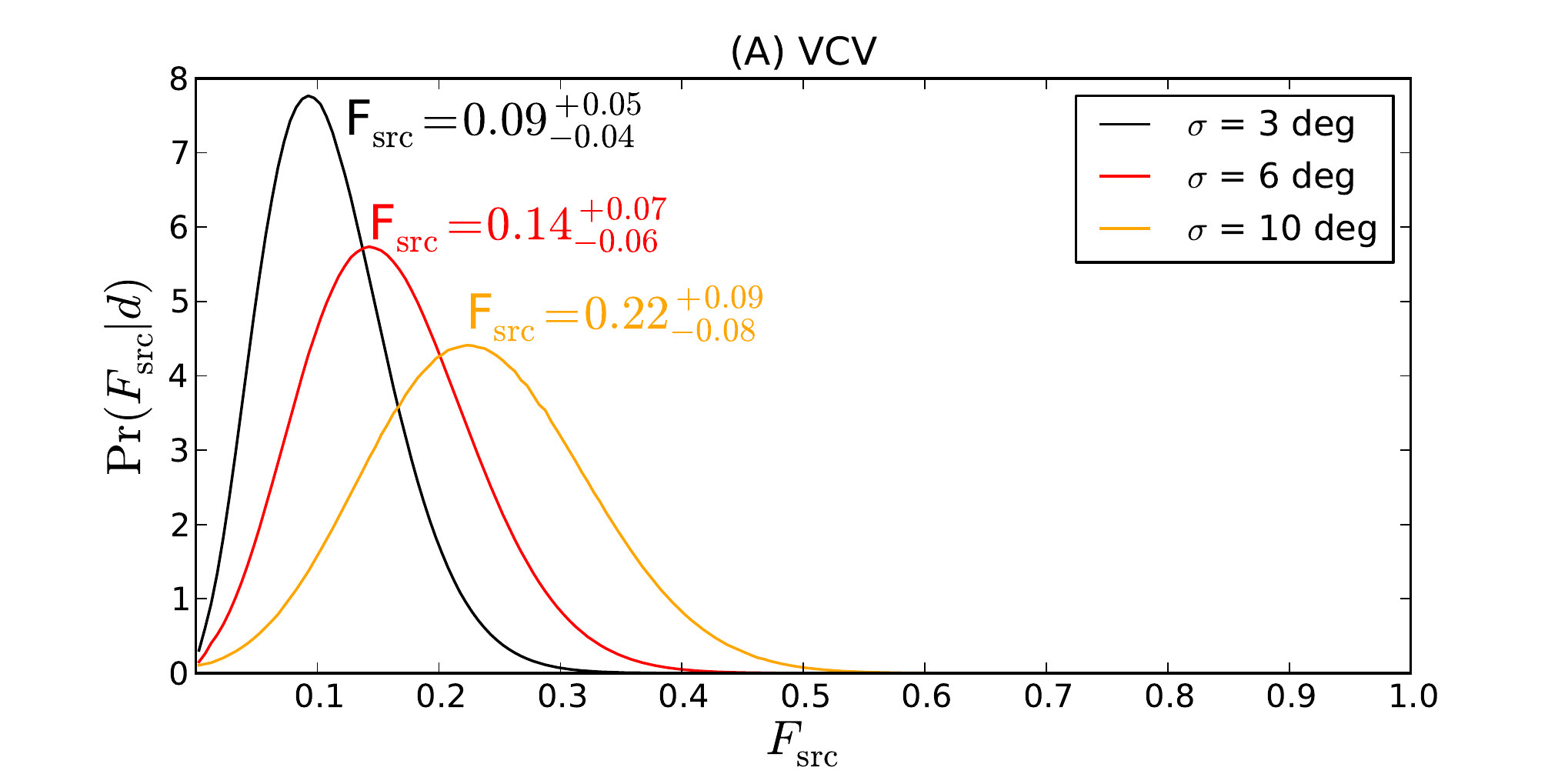}&\includegraphics[width=85mm]{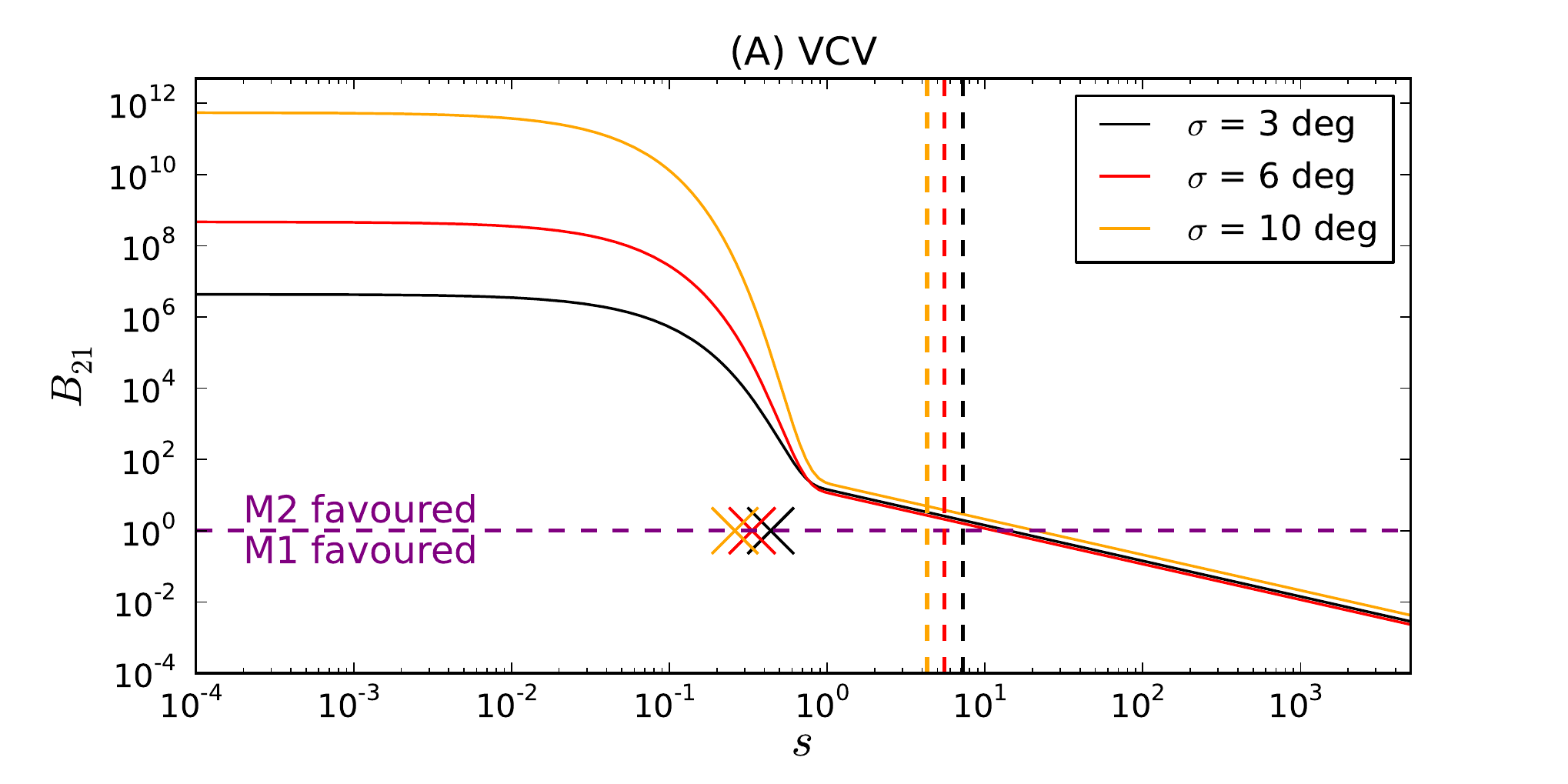}\\
\end{array}$
\end{center}
\vspace{-0.5cm}
\end{figure*}

\begin{figure*}
\begin{center}$
\arraycolsep=0.01pt\def\arraystretch{0.01}
\begin{array}{cc}
\includegraphics[width=85mm]{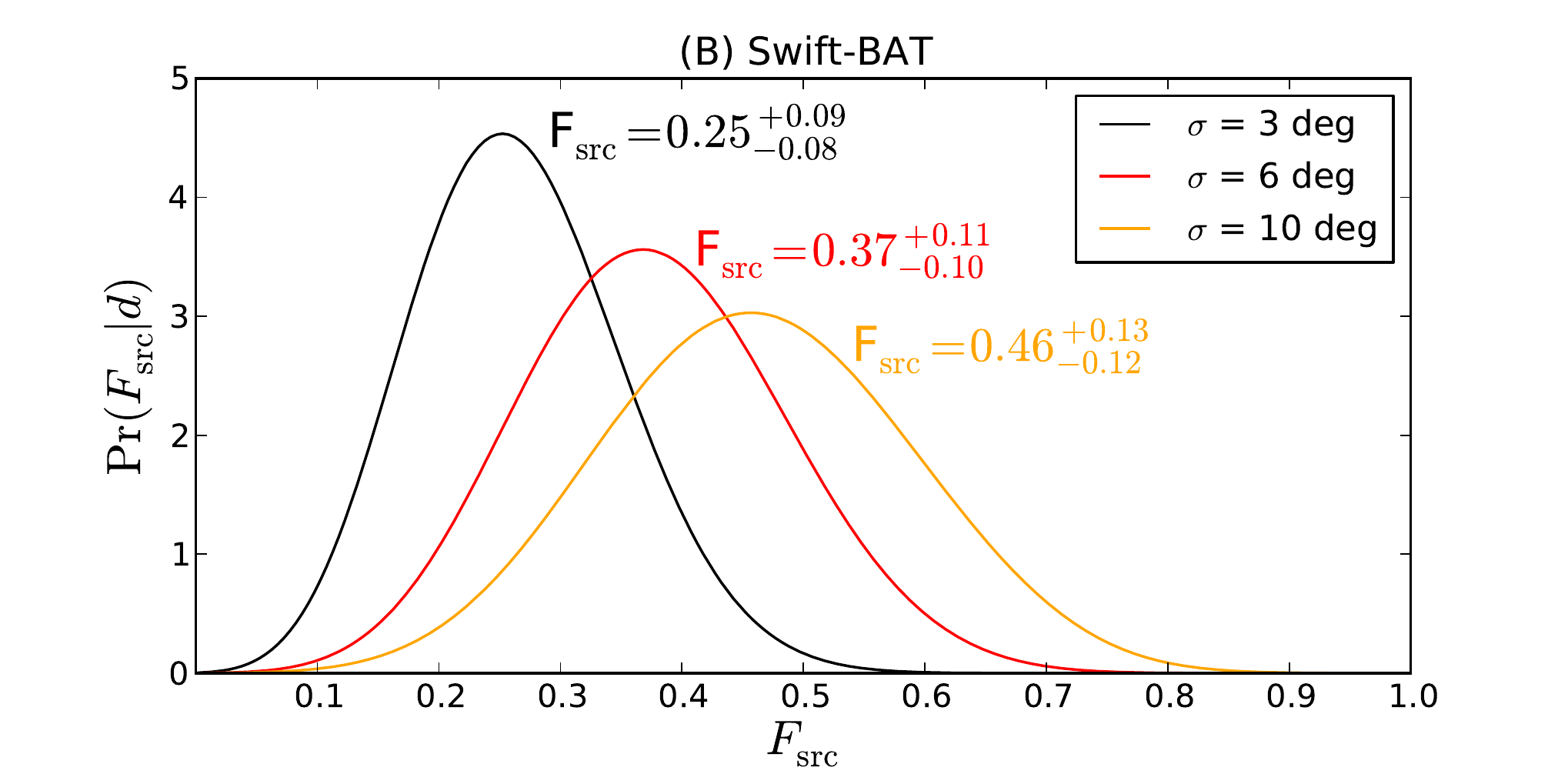}&\includegraphics[width=85mm]{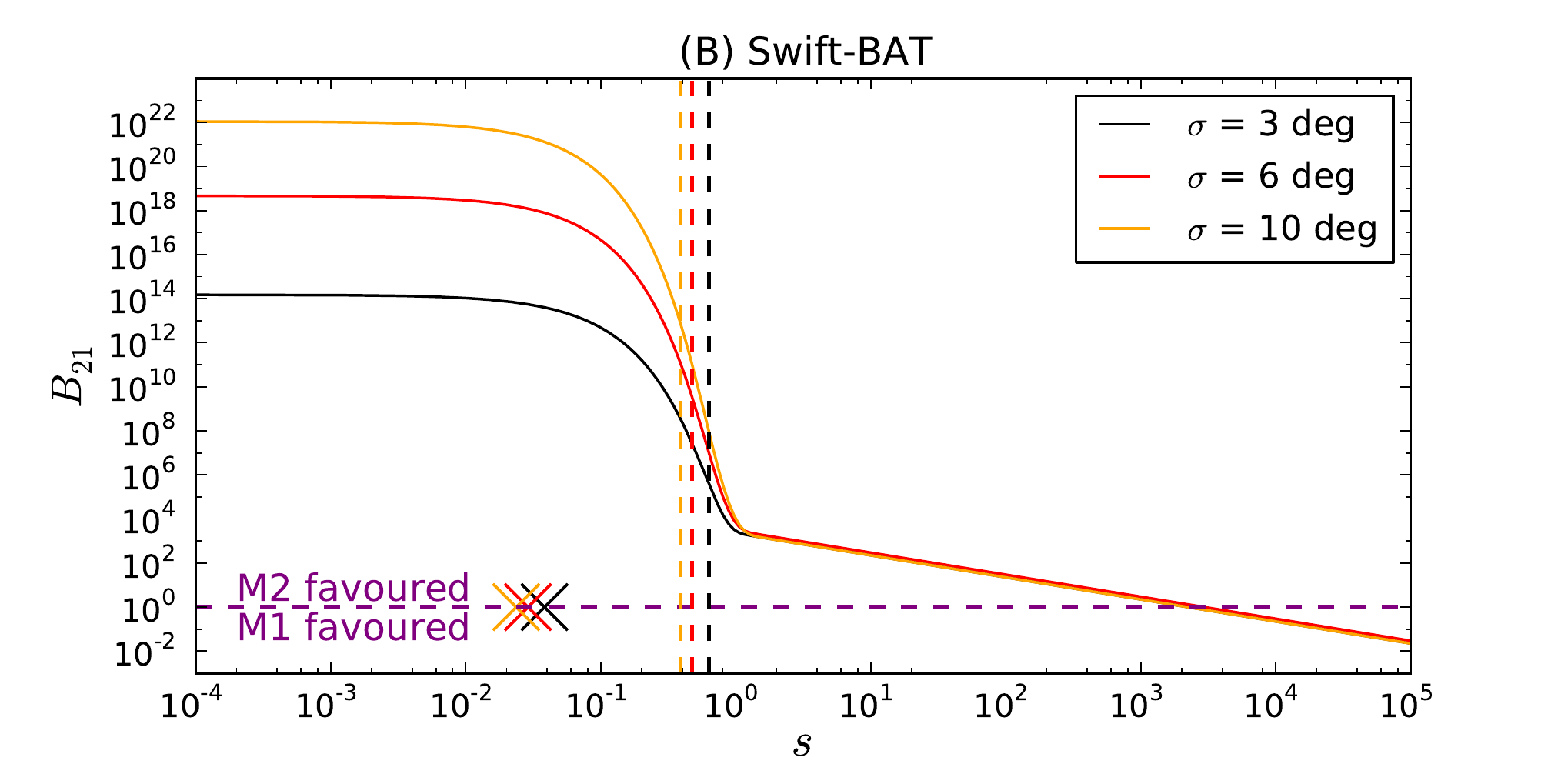}\\
\end{array}$
\end{center}
\vspace{-0.5cm}
\end{figure*}

\begin{figure*}
\begin{center}$
\arraycolsep=0.01pt\def\arraystretch{0.01}
\begin{array}{cc}
\includegraphics[width=85mm]{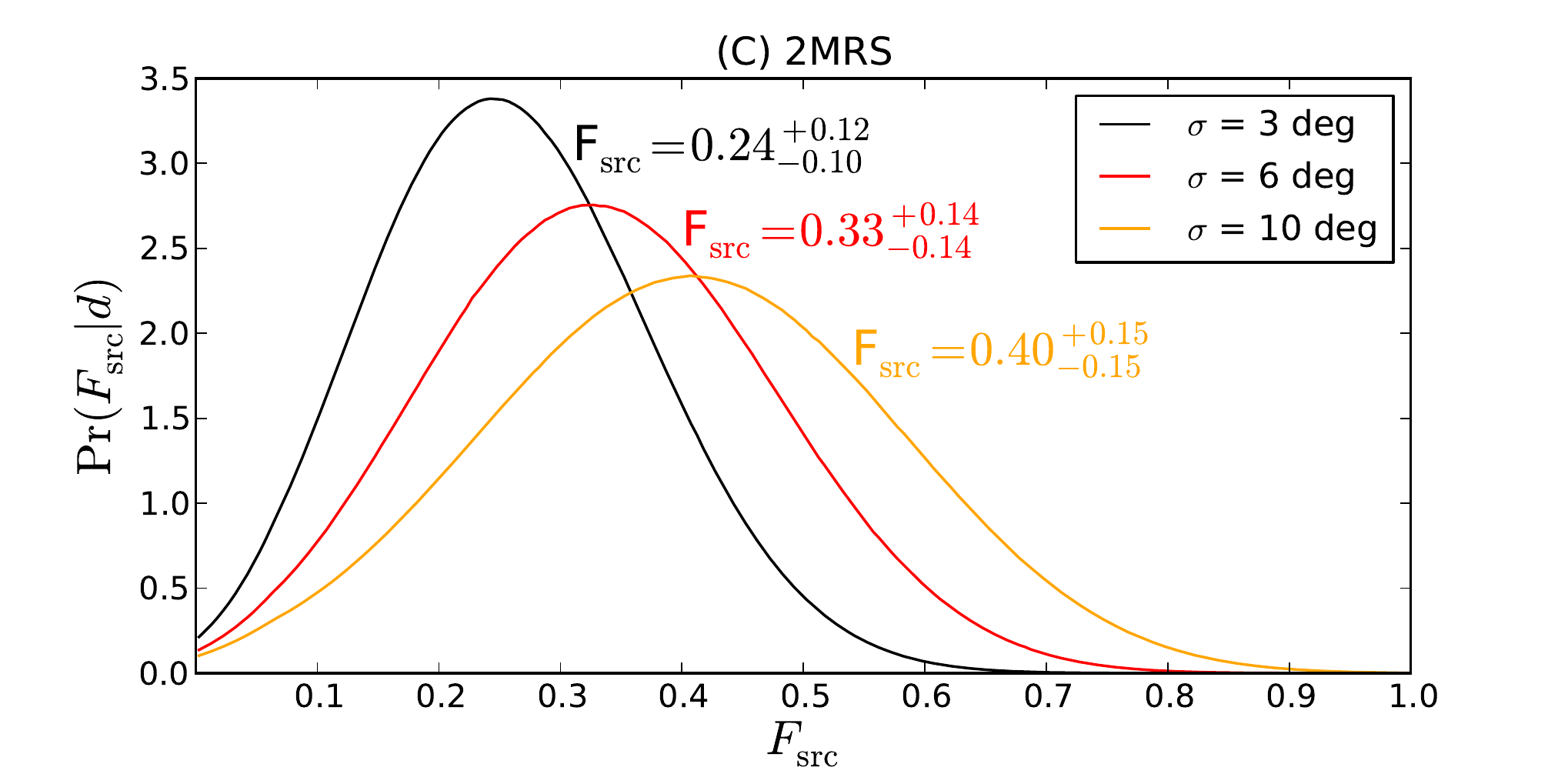}&\includegraphics[width=85mm]{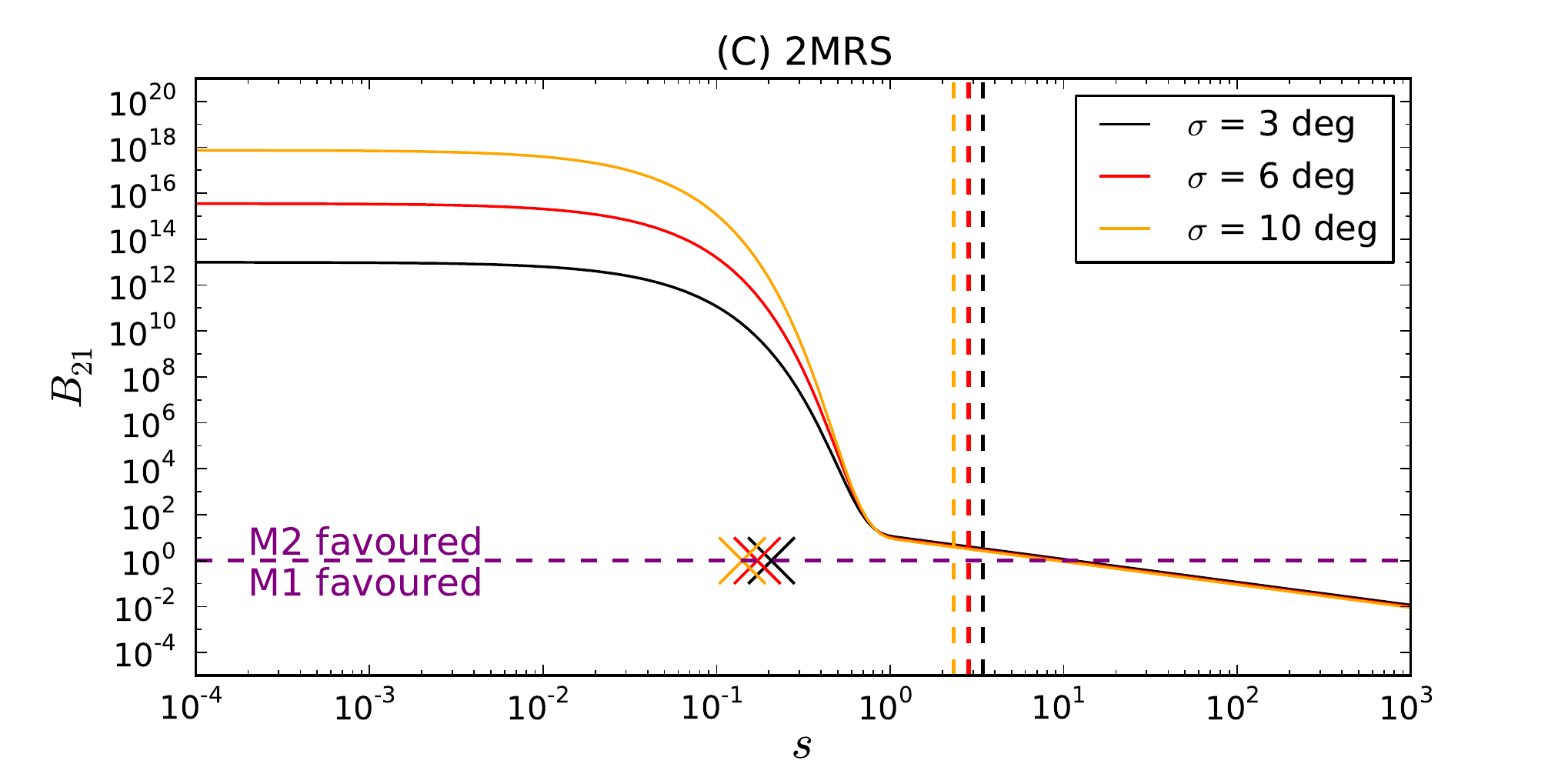}\\
\end{array}$
\end{center}
\label{fig:ThreeCats}
\vspace{-0.5cm}
\end{figure*}

\begin{figure*}
\begin{center}$
\arraycolsep=0.01pt\def\arraystretch{0.01}
\begin{array}{cc}
\includegraphics[width=85mm]{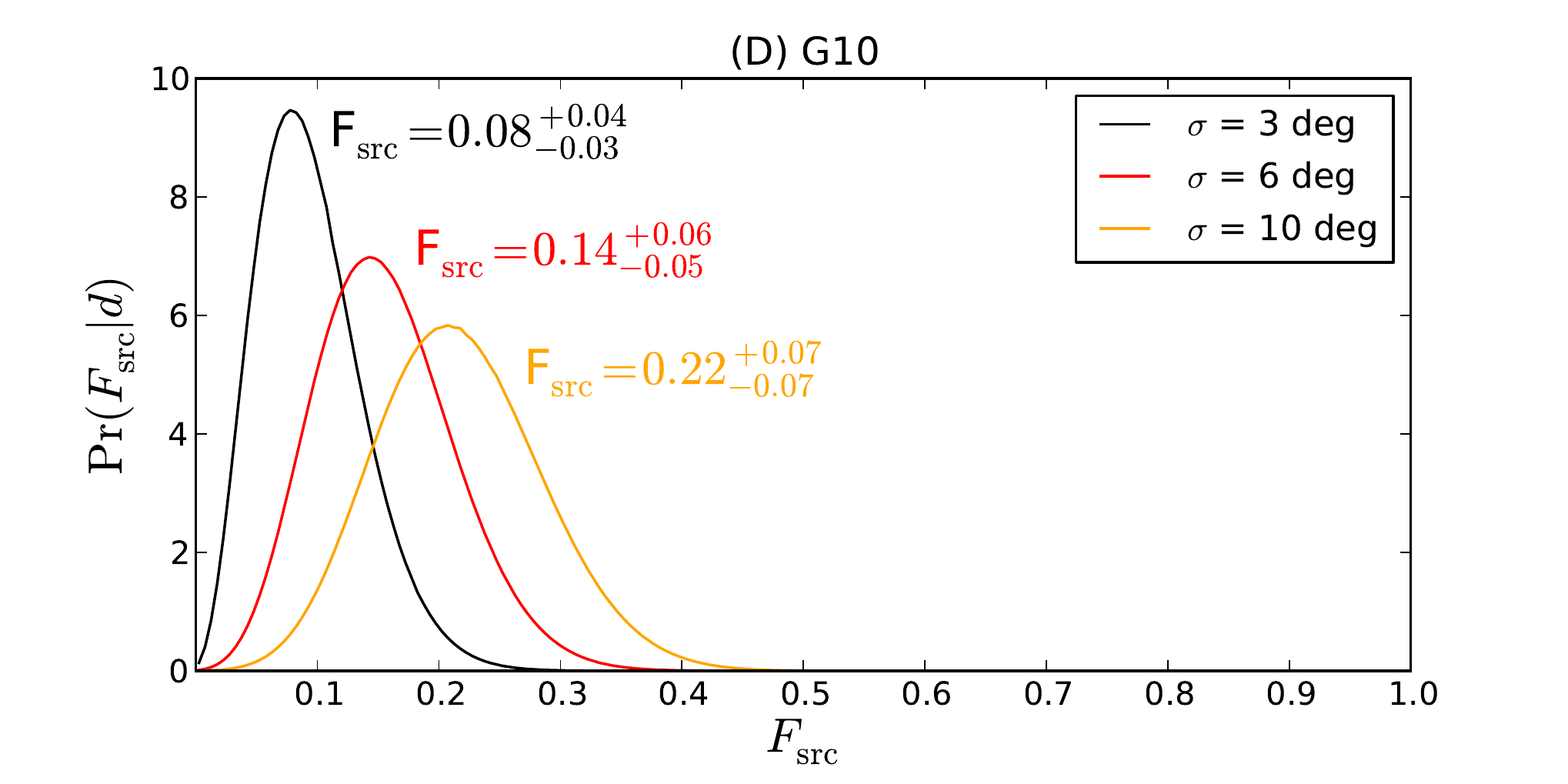}&\includegraphics[width=85mm]{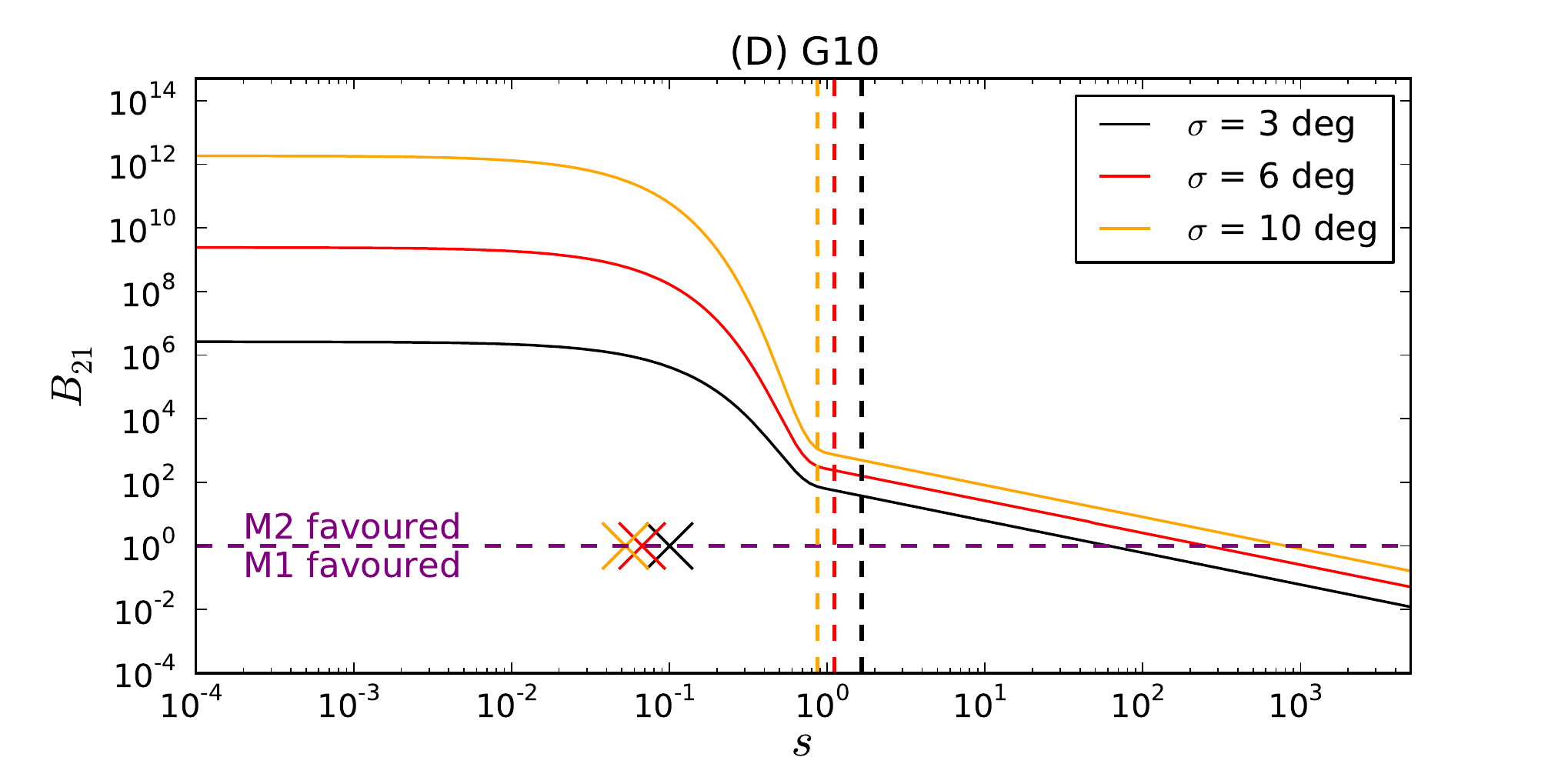}\\
\end{array}$
\end{center}
\caption{Posteriors of the source fraction, and plots of $B_{21}$ against the hyperparameter $s$, for the three smearing angles $\sigma=3 \,\, \rm{deg}$, $6\, \, \rm{deg}$ and $10 \,\, \rm{deg}$, and for the three source catalogs (A) VCV, (B) Swift-BAT, and (C) 2MRS. The plots of $B_{21}$ show physically plausible ranges of $s$: The $\times$-marks and the vertical lines signify the minimum and the maximum values of these ranges.
}
\label{fig:ThreeCats}
\end{figure*}
\twocolumn

\newpage


 



\newpage
\onecolumn

\appendix

\section{Numerical evaluation of the likelihood}
\label{sec:LikeApp}

The likelihood, as given in Equation~\ref{eq:TotalLikelihood}, is a product over Poisson likelihoods for the individual pixels,
\begin{equation}
\label{eq:TotalLikelihoodApp}
\rm{Pr(} { \textbf{\emph{d}} | \Gamma,\emph{R} }\rm{)} = \prod^{\emph{N}_{\rm{p}}}_{\emph{p}=1} \frac{( \overline{\emph{N}}_{\rm{src},\emph{p}} + \overline{\emph{N}}_{\rm{bkg},\emph{p}} )^{\emph{N}_{\rm{c},p}} \exp[-( \overline{\emph{N}}_{\rm{src},\emph{p}} + \overline{\emph{N}}_{\rm{bkg},\emph{p}} )]  }{\emph{N}_{\rm{c},p}!}    ,
\end{equation} 
where the product is over the pixels, $\emph{N}_{\rm{c},\emph{p}}$ is the number of counts in pixel $\emph{p}$, and $\overline{\emph{N}}_{\rm{bkg},\emph{p}}$ and $\overline{\emph{N}}_{\rm{src},\emph{p}}$ are the expected numbers of counts from the background and sources in pixel $p$. This expression for the likelihood proved to be inefficient for use, as it required a great number of computations: The total number of pixels was $\emph{N}_{\rm{p}} = 1800 \times 3600 = \,$ 6,480,000. If a $\Gamma \times \emph{R}$ grid of $100 \times 100$ is used, a total of 64,800,000,000 calculations would be required. 
\par
The total number of calculations can be greatly reduced by rearranging the expression. For a given data set, we can separate the product of Equation~\ref{eq:TotalLikelihoodApp} into a product over those pixels that include an event, $\{q\}$, and pixels that do not, $\{r\}$. Using the fact that $N_{q}=1$ for all $\{q\}$ and $N_{r}=0$ for all $\{r\}$, we can write 
\begin{equation}
\label{eq:BayesTheorem0}
\rm{Pr(} { \textbf{\emph{d}} | \Gamma,\emph{R} }\rm{)}  =  \prod^{\emph{N}_{r}}_{\emph{r}=1} \exp[-( \overline{\emph{N}}_{\rm{src},\emph{r}} + \overline{\emph{N}}_{\rm{bkg},\emph{r}})] \times
\prod^{\emph{N}_{q}}_{\emph{q}=1} ( \overline{\emph{N}}_{\rm{src},\emph{q}} + \overline{\emph{N}}_{\rm{bkg},\emph{q}} ) \exp[-( \overline{\emph{N}}_{\rm{src},\emph{q}} + \overline{\emph{N}}_{\rm{bkg},\emph{q}})]       
\end{equation} 
\begin{equation}
\label{eq:BayesTheorem}
\hspace{1.95cm} =  \exp[-(  \Gamma \Sigma_{\rm{src}} + \emph{R} \Sigma_{\rm{bkg}}  )] \times
\prod^{\emph{N}_{\rm{q}}}_{\emph{q}=1} ( \overline{\emph{N}}_{\rm{src},\emph{q}} +  \overline{\emph{N}}_{\rm{bkg},\emph{q}}) \exp[-( \overline{\emph{N}}_{\rm{src},\emph{q}} + \overline{\emph{N}}_{\rm{bkg},\emph{q}}  )] .
\label{eq:fromtobe}
\end{equation} 
where $\Sigma_{\rm{src}}=\sum^{\emph{N}_{\rm{r}}}_{\emph{r}=1}\emph{m}_{\rm{src},\emph{r}}$ and $\Sigma_{\rm{bkg}}=\sum^{\emph{N}_{\rm{r}}}_{\emph{r}=1}\emph{m}_{\rm{bkg},\emph{r}}$, and $\emph{m}_{\rm{src},\emph{p}}$ and $\emph{m}_{\rm{bkg},\emph{p}}$ are two pixelized maps obeying the equations
\begin{equation}
\overline{\emph{N}}_{\rm{src},\emph{p}} =   \Gamma m_{ \rm{src},\emph{p} } . 
\end{equation} 
\begin{equation}
\overline{\emph{N}}_{\rm{bkg},\emph{p}} =   \emph{R}m_{ \rm{bkg},\emph{p} } 
\end{equation} 
Thus, the initial expression has been rearranged in such a way that the vast majority of Poisson calculations is contained within the sums $\Sigma_{\rm{src}}$ and $\Sigma_{\rm{bkg}}$. These sums can be calculated in advance for the entire grid of $\Gamma$ and $\emph{R}$. This greatly reduces the total number of calculations required for Equation~\ref{eq:TotalLikelihoodApp}, and speeds up the full calculation by a factor of $\sim 10^5$.

\section{Model comparison and prior sensitivity}
\label{sec:BayesApp}

The Bayes factor that was discussed in Section~\ref{section:BF} is comparing two models: A simple model $M_{1}$ of uniform UHECRs, and a more complex model $M_{2}$ that has both uniform and sourced UHECRs. As explained in the section, due to $M_{1}$ being nested within $M_{2}$, the expression for the Bayes factor reduces to
\begin{equation}
\label{eq:nochmalrateApp}
B_{12} = \frac { \int {\rm Pr}(\Gamma = 0, R|\textbf{\emph{d}}, M_2) \, \rm{d}\emph{R} }{ \int {\rm Pr}(\Gamma = 0, R| M_2) \, \rm{d}\emph{R} } = \frac {  {\rm Pr}(\Gamma = 0 |\textbf{\emph{d}}, M_2) }{  {\rm Pr}(\Gamma = 0| M_2) }, \\
\label{eq:SDDRApp}
\end{equation}
where $\Gamma$  and $R$ are the background and source rates and $\textbf{\emph{d}}$ are the data.
Qualitatively, the expression means that the nested uniform model is preferred if, within the context of the more complex model, the data result in an increased probability that $\Gamma = 0$. A uniform prior was used, given by 
\begin{equation}
\label{eq:priorprior1}
{\rm Pr}(\Gamma, R| M_2) = \frac{ 1 }{ s^2 \rm{\Gamma_{max}} \emph{R}_{\rm{max}} },
\end{equation} 
where $s$ is the hyperparameter that determines the width of the prior. $\rm{\Gamma_{max}}$ and $R_{\rm{max}}$ have been chosen in such a way that when $s=1$, the prior covers the 99.7\% credible region implied by the likelihood and an infinitely broad uniform prior. To explain the dependence of the Bayes factor on $s$, three illustrative cases are used: The case of a simple Gaussian likelihood, the case of the Poisson product likelihood of Equation~\ref{eq:TotalLikelihood}, and the likelihood of On/Off measurements.

\subsection{Gaussian likelihood}

We consider the case of a Gaussian likelihood given by
\begin{equation}
\label{eq:nochmalrate}
\rm{Pr(}  \textbf{\emph{d}}  |   \Gamma,\emph{R}  \rm{)} = \frac{1}{2\pi \rm{\sigma_\Gamma}  \rm{\sigma_\emph{R}} } \exp\bigg[-\frac{ (\Gamma-\rm{\Gamma_{\rm{\mu}}})^2}{2\rm{\sigma_\Gamma}^2}  \bigg]\bigg[-\frac{(\emph{R}-\emph{R}_{\rm{\mu}})^2}{2\rm{\sigma_\emph{R}}^2}  \bigg],
\end{equation}
where $\rm{\Gamma_{\rm{\mu}}}$ and $\emph{R}_{\rm{\mu}}$ are the coordinates of the likelihood mean, $\rm{\sigma_\Gamma}$ and $\rm{\sigma_\emph{R}}$ are the standard deviations on the two parameters.

This likelihood is shown in the upper panel of Figure~\ref{fig:GaussNInt}, focusing on three regions $s=0.1, \, 1, \, 2$. These regions correspond to the regions over which the flat prior is taken for these values of the hyperparameter. The lower panel shows the posteriors $\rm{Pr}(\Gamma, R|\textbf{\emph{d}}, M_2)$ for the same $s$ values. As the priors are flat, the posteriors are equivalent to the likelihood in the prior region, normalized over the prior region.
\par
These posteriors can be used to illustrate the dependence of the Bayes factor in Equation~\ref{eq:SDDRApp} on the hyperparameter $s$. For $s>1$, the numerator ${\rm Pr}(\Gamma = 0 |\textbf{\emph{d}}, M_2)$ is constant, as $\rm{Pr}(\Gamma, \emph{R}|\textbf{\emph{d}}, M_2)$ corresponds to the normalized likelihood, and does not vary as $s$ is increased beyond $s=1$. The denominator ${\rm Pr}(\Gamma = 0| M_2)$ falls linearly with $s$. Thus, we expect that for $s>1$, $B_{12}$ increases linearly with $s$. 
\par
For lower values of $s$, the behaviour of $B_{12}$ is more complicated, as can be seen in the left-hand lower panel of Figure~\ref{fig:GaussNInt}. For low values of $s$, the likelihood becomes 
\begin{equation}
\label{eq:nochmalrate}
\rm{Pr(} \Gamma, \emph{R} | \textbf{\emph{d}} \rm{)} = \frac{1}{2\pi  \rm{\sigma_\Gamma} \rm{\sigma_\emph{R}} } e^{\frac{ -\rm{\Gamma_{\rm{\mu}}}^2}{2\rm{\sigma_\Gamma}^2}} e^{\frac{ -\emph{R}_{\rm{\mu}}^2}{2\rm{\sigma_\emph{R}}^2}}  \Big(1 + \frac{ \rm{\Gamma_{\rm{\mu}}} \Gamma}{2\rm{\sigma_\Gamma}^2} + \frac{ \emph{R}_{\rm{\mu}} \emph{R}}{2\rm{\sigma_\emph{R}}^2} \Big) .
\end{equation}
This means that the posterior becomes linear and increasingly flat as $s \rightarrow 0$. As the function becomes increasingly flat, the ratio in Equation~\ref{eq:SDDRApp} becomes a ratio of two normalized flat functions, so that qualitatively, we can expect it to approach unity. This can also be shown more rigorously, as for low values of $\Gamma$ and $R$, Equation~\ref{eq:SDDRApp} reduces to
\begin{equation}
\label{eq:nochmalrate}
B_{12}  =   1 - s\frac{\Gamma_\mu \Gamma_{\rm{max}} }{2{\sigma_\Gamma}^2} .
\end{equation}

\begin{figure*}
  \centering $
\arraycolsep=0.01pt\def\arraystretch{-0.5}
\begin{array}{ccc}
\includegraphics[width=85mm]{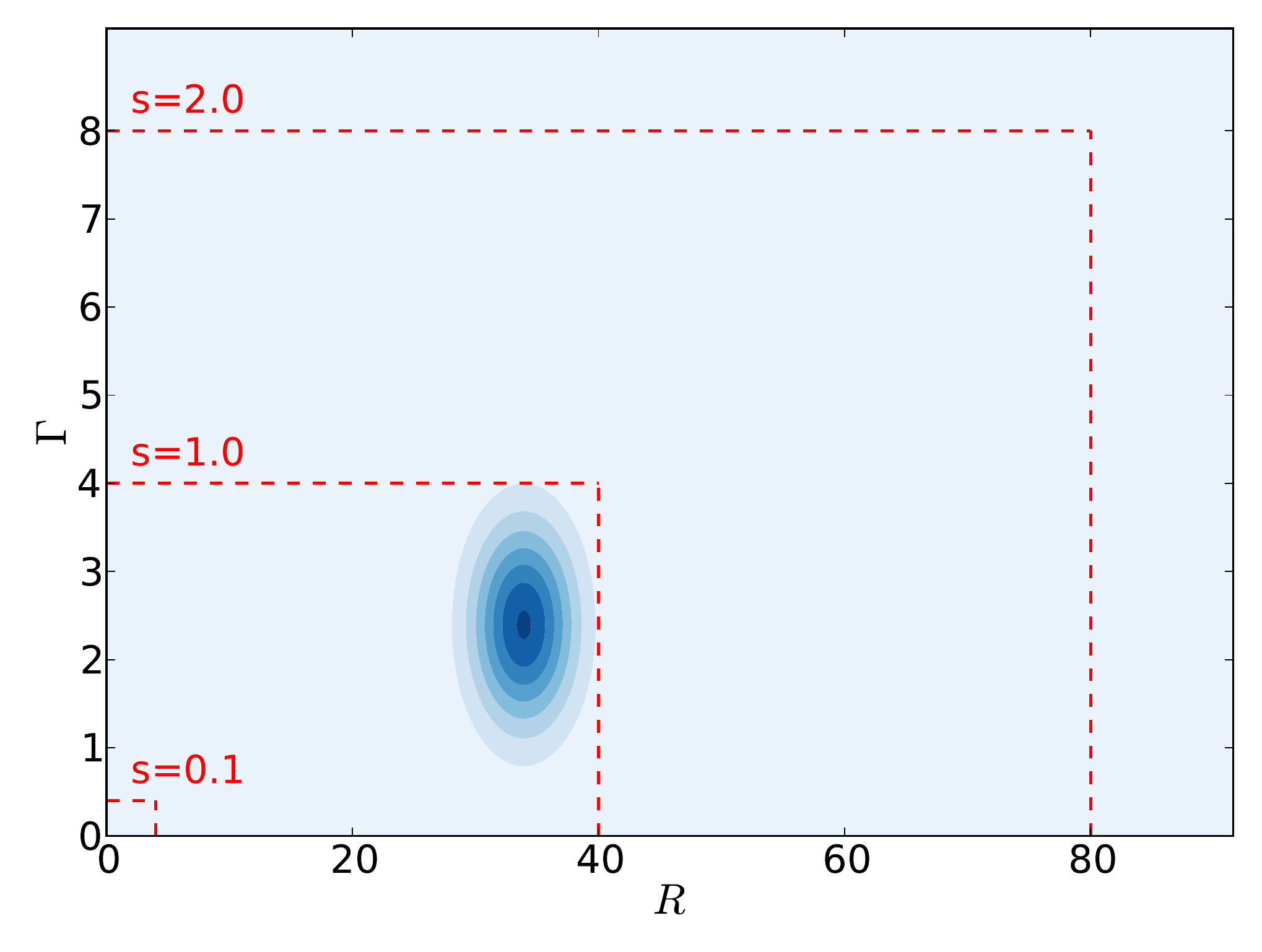}&
\end{array}$
\label{fig:GaussNInt}
\end{figure*}

\begin{figure}
\begin{center}$
\arraycolsep=0.000000000000000000000pt
\begin{array}{ccc}
\hbox{\includegraphics[width=62mm]{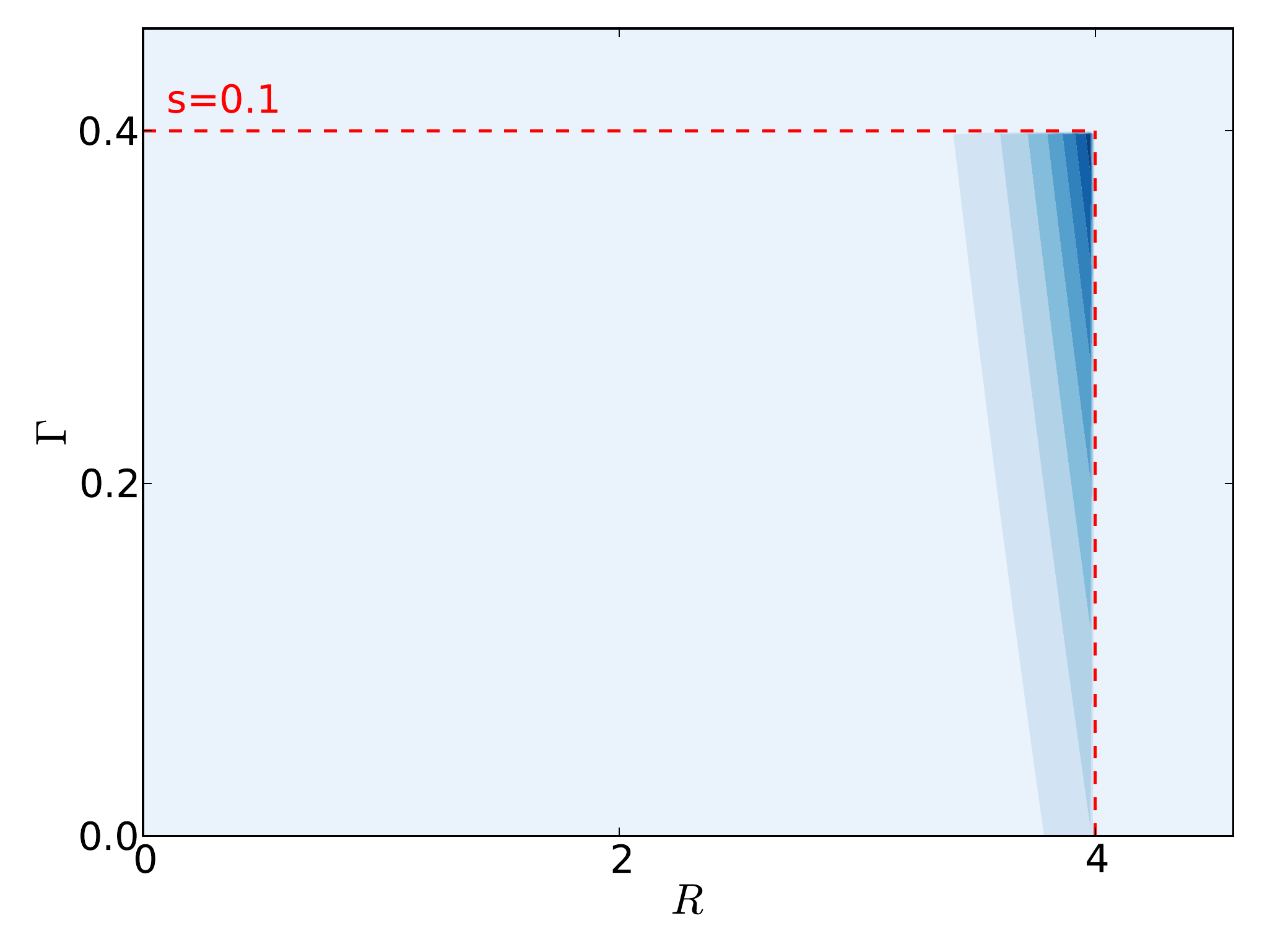}} & \includegraphics[width=62mm]{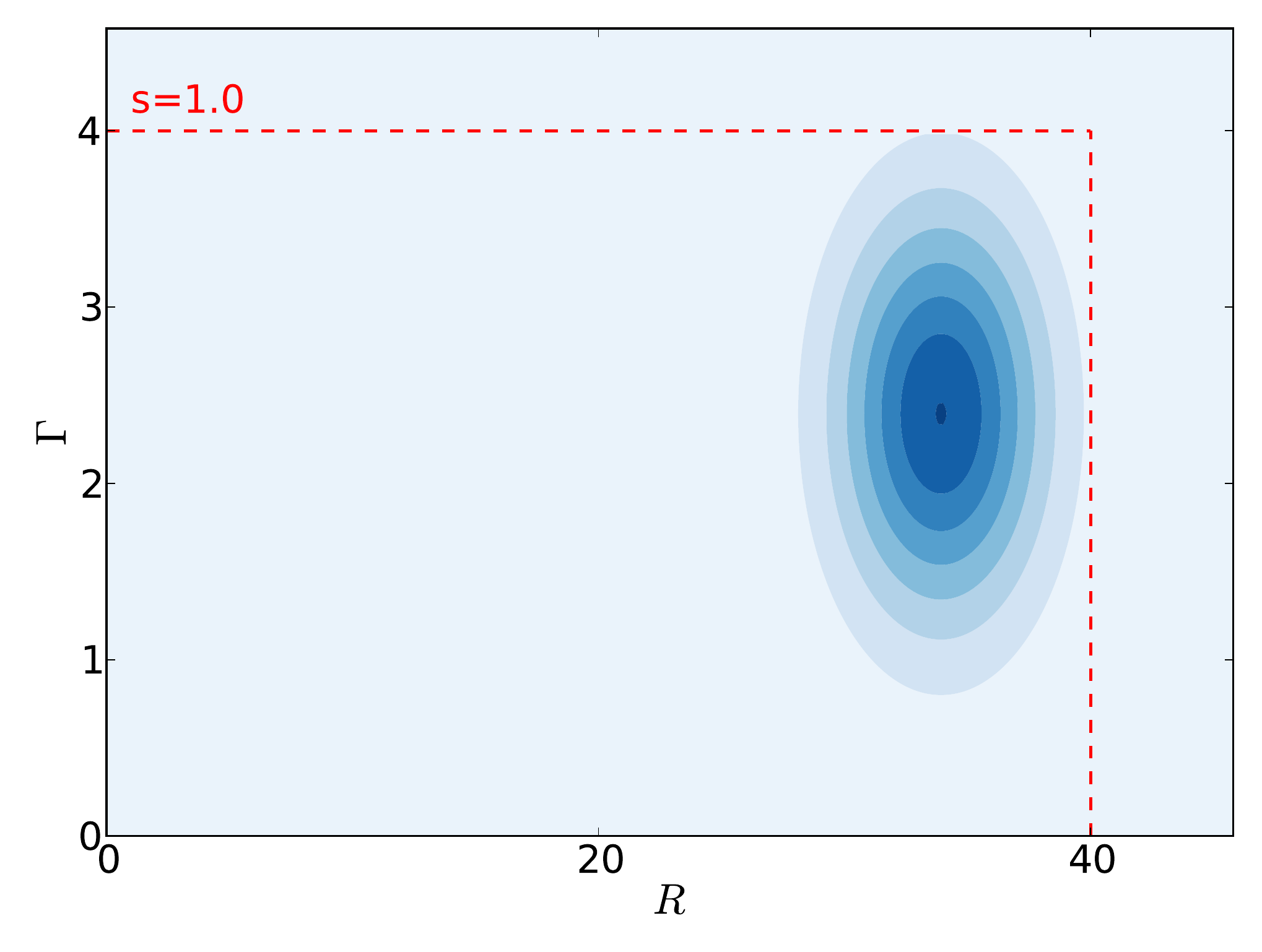} & \includegraphics[width=62mm]{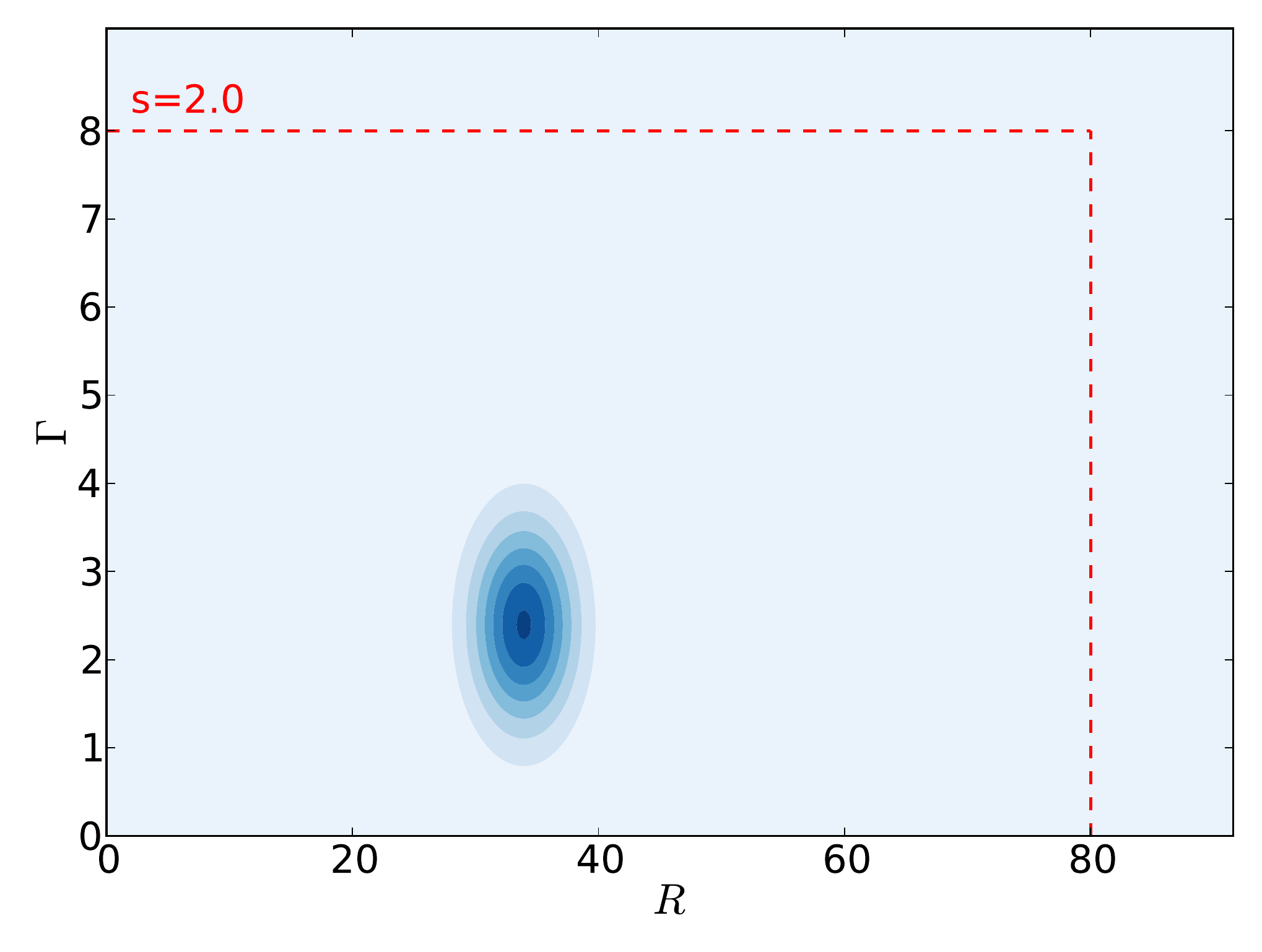} \\
\end{array}$
\end{center}
  \caption{Upper panel: Example Gaussian likelihood. The red lines denote prior regions for three different values of the hyperparameter $s$. Lower panel: Posteriors for the same $s$ values are displayed. }
\label{fig:GaussNInt}
\end{figure}

\begin{figure*}
  \centering $
\arraycolsep=0.01pt\def\arraystretch{-0.5}
\begin{array}{ccc}
\includegraphics[width=85mm]{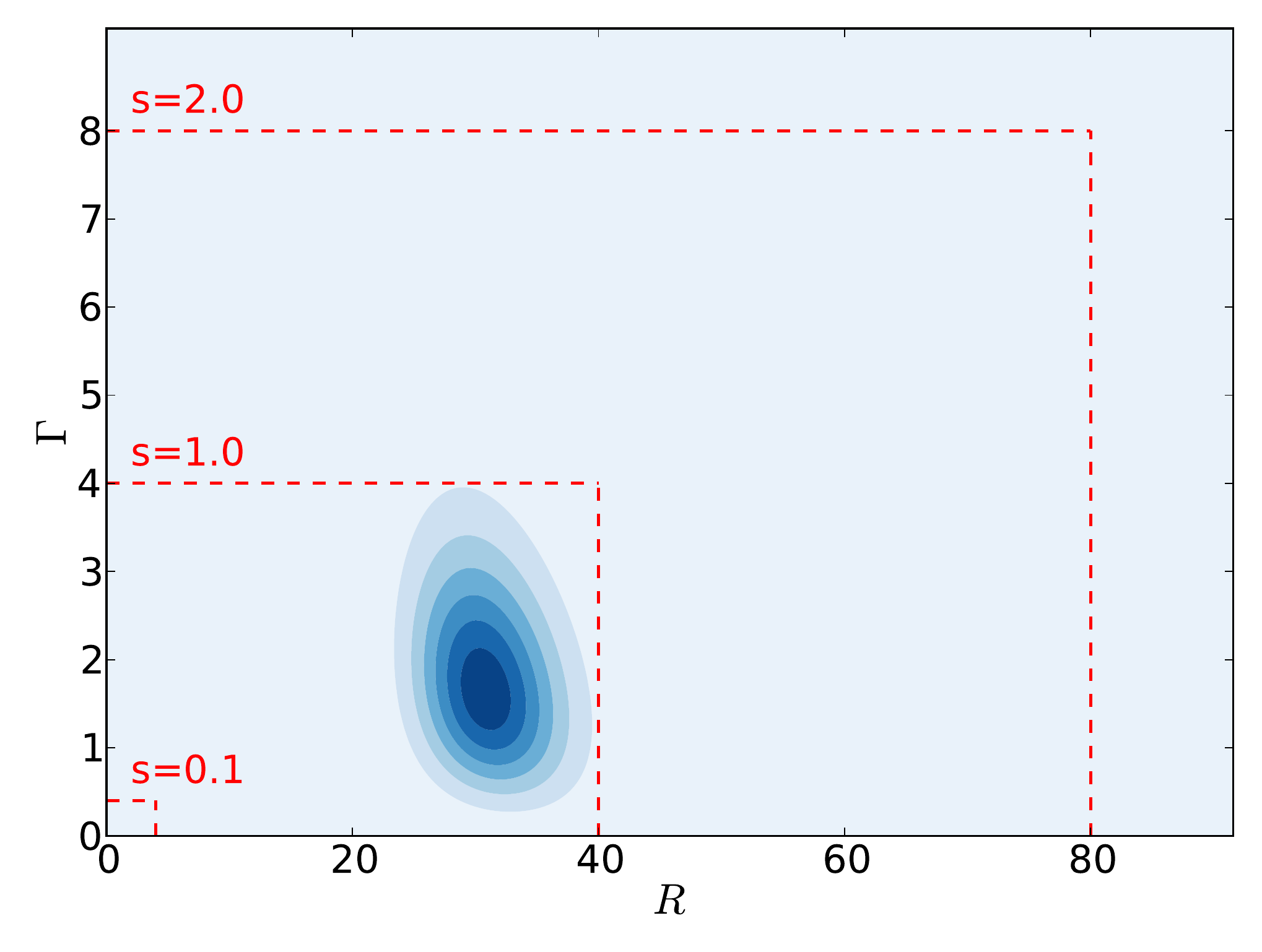}&
\end{array}$
\label{fig:PoissNInt0}
\end{figure*}

\begin{figure}
\begin{center}$
\arraycolsep=0.000000000000000000000pt
\begin{array}{ccc}
\hbox{\includegraphics[width=62mm]{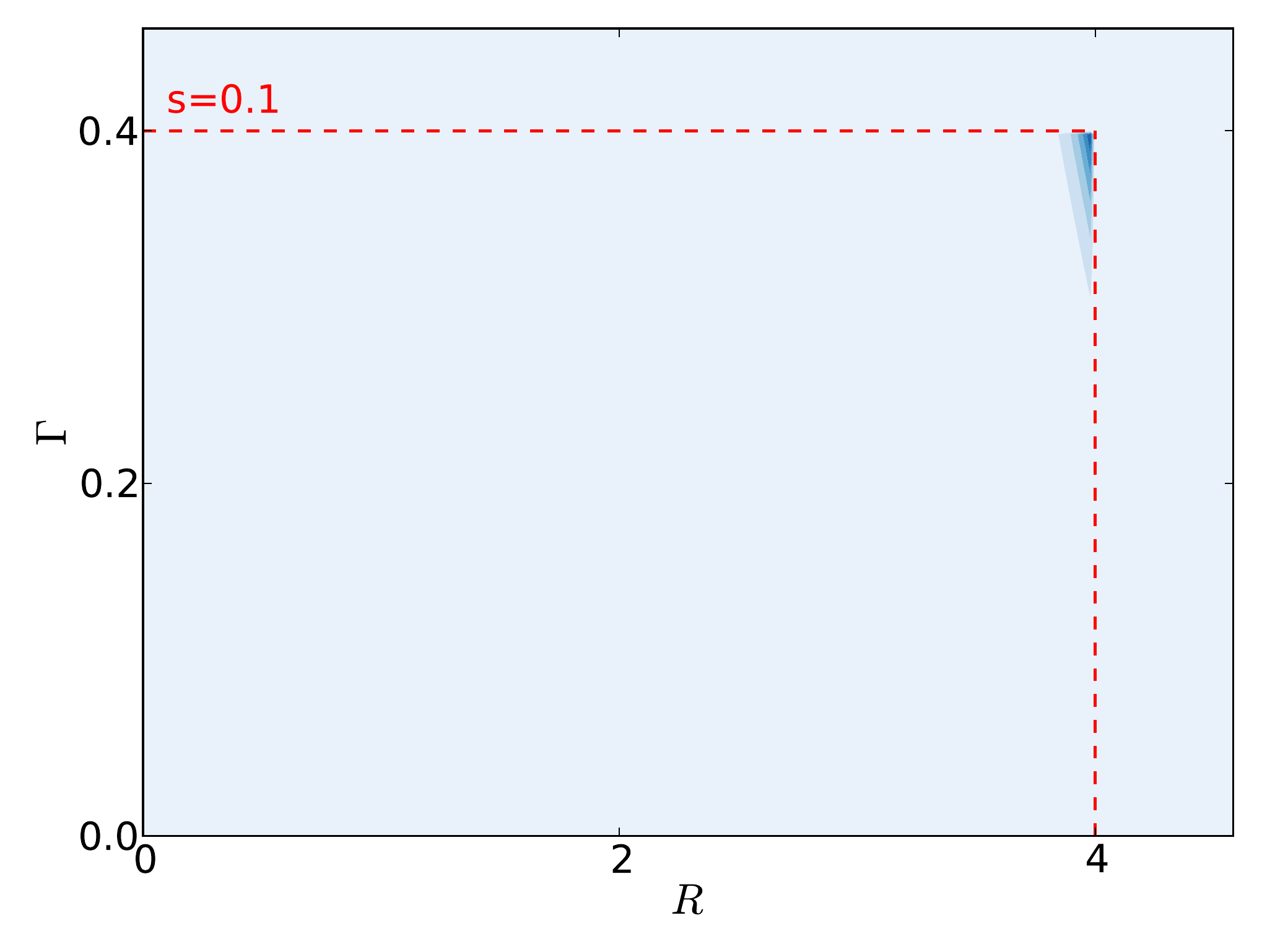}} & \includegraphics[width=62mm]{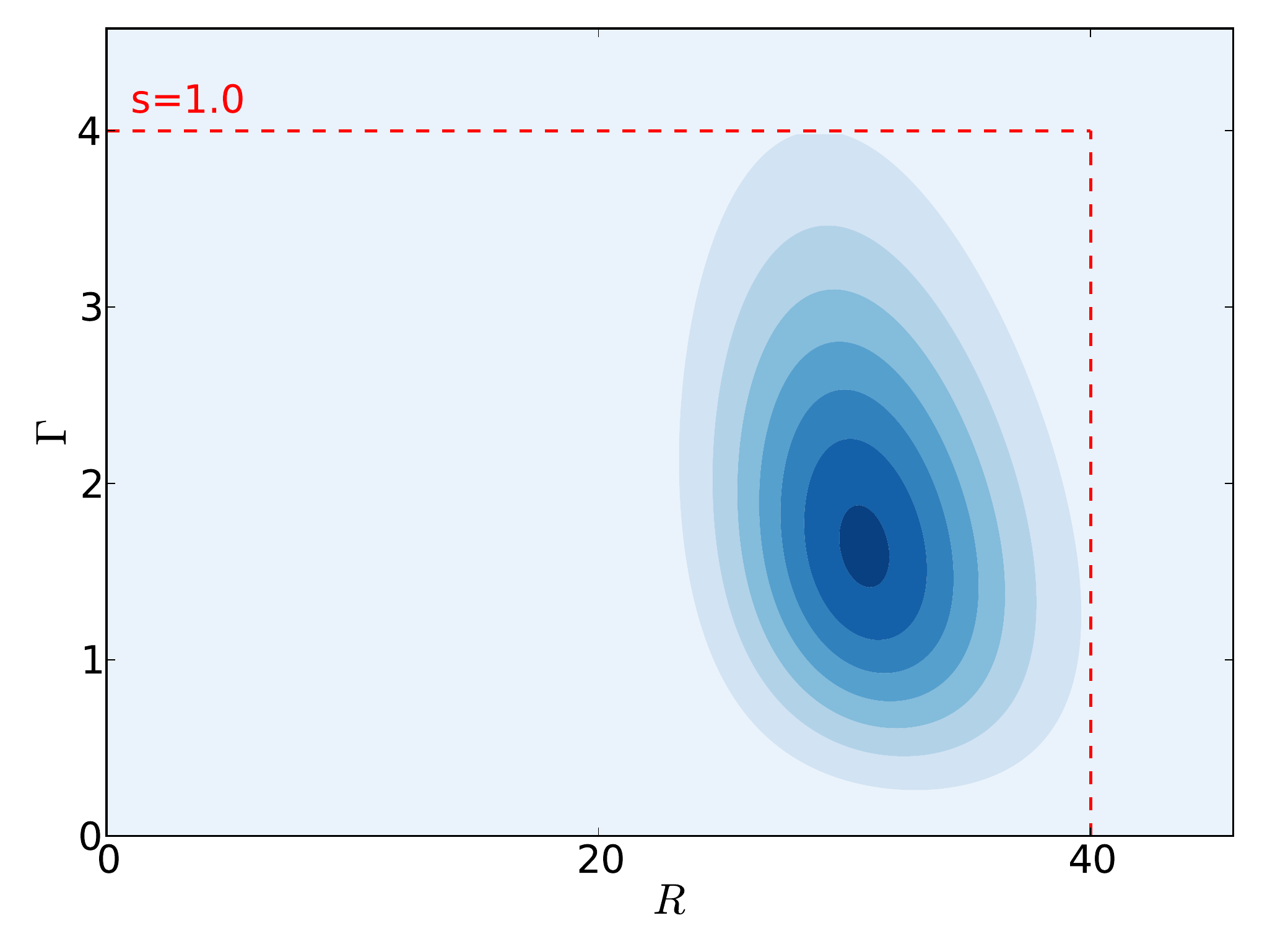} & \includegraphics[width=62mm]{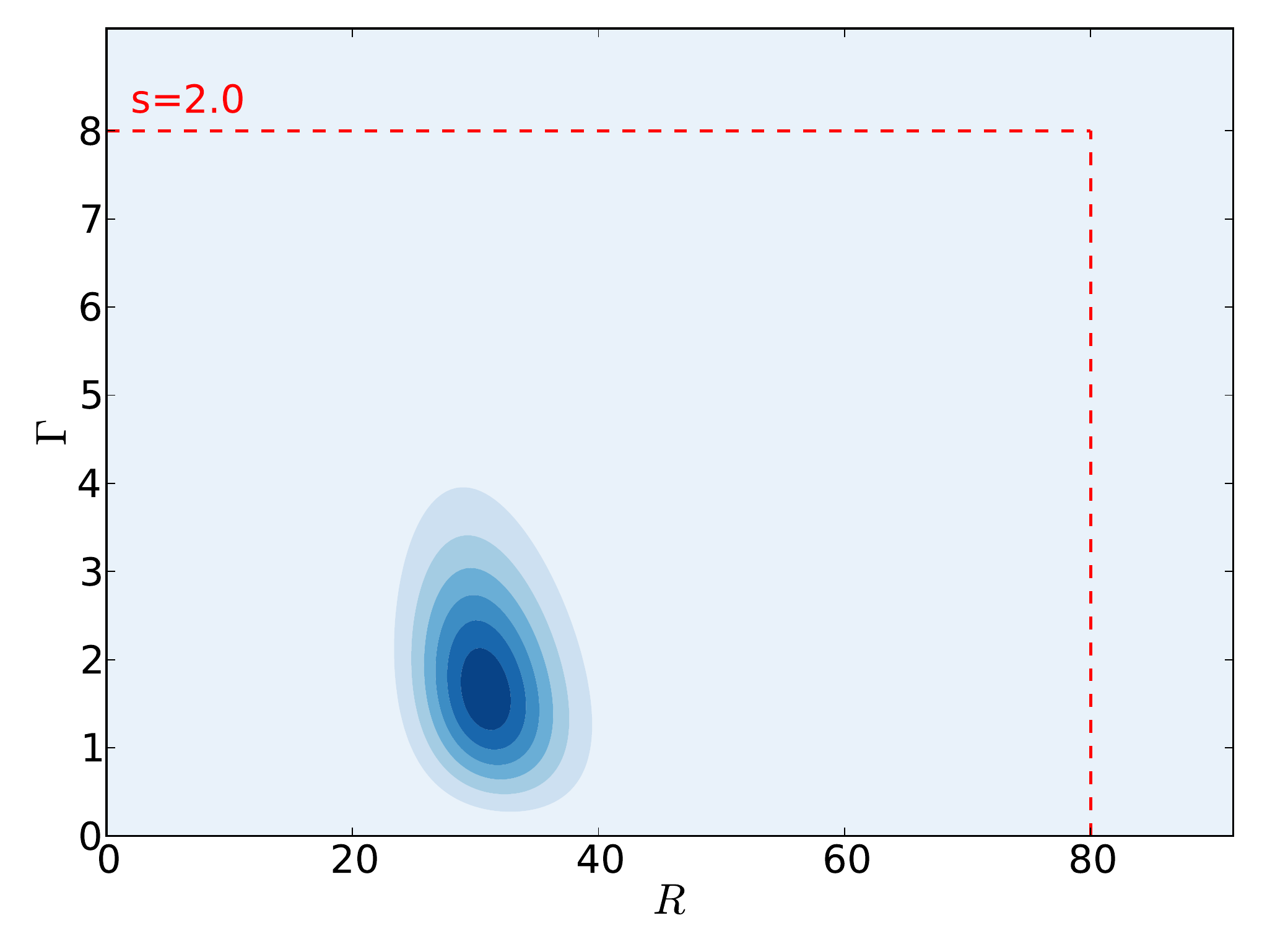} \\
\end{array}$
\end{center}
\caption{Upper panel: Example Poisson product likelihood. The red lines denote prior regions for three different values of the hyperparameter $s$. Lower panel: Posteriors for the same $s$ values are displayed. }
\label{fig:PoissNInt}
\end{figure}

\subsection{Poisson product likelihood}

We consider the same likelihood that was used in Equation~\ref{eq:TotalLikelihood}. The total likelihood is a product of individual Poisson likelihoods for 6,480,000 pixels, and can be written as

\begin{equation}
\label{eq:TotalLikelihoodPoiss}
\rm{Pr(} { \textbf{\emph{d}} | \Gamma,\emph{R} }\rm{)} = \prod^{\emph{N}_{\rm{p}}}_{\emph{p}=1} \frac{(  \overline{\emph{N}}_{\rm{src},\emph{p}}+\overline{\emph{N}}_{\rm{bkg},\emph{p}})^{\emph{N}_{\rm{c},\emph{p}}} \exp[-(\overline{\emph{N}}_{\rm{src},\emph{p}}+ \overline{\emph{N}}_{\rm{bkg},\emph{p}} )]  }{\emph{N}_{\rm{c},\emph{p}}!}     ,
\end{equation}
where $\overline{\emph{N}}_{\rm{src},\emph{p}}$ and $\overline{\emph{N}}_{\rm{bkg},\emph{p}}$ are the expected numbers of counts in pixel $\emph{p}$ due to source and background rates, respectively. Figure~\ref{fig:PoissNInt} shows the likelihood, as well as three prior regions, and the posteriors calculated for these three regions.
\par
For high values of $s$, the posterior looks very much like the Gaussian, so that we expect the same behaviour for the Bayes factor, including the linear behaviour for $s>1$. A difference arises at small values of $s$. Here, we see that most of the posterior is concentrated at the highest values of $\Gamma$ and $R$. As the product of Equation~\ref{eq:TotalLikelihoodPoiss}, for low values of $\Gamma$ and $R$, reduces to
\begin{equation}
\label{eq:LALALA}
\rm{Pr(} { \textbf{\emph{d}} | \Gamma,\emph{R} }\rm{)} = \prod^{\emph{N}_{\rm{q}}}_{\emph{q}=1} (  \Gamma \emph{m}_{ \rm{src},\emph{q} } + \emph{R}\emph{m}_{ \rm{bkg},\emph{q} })    ,
\end{equation}
where $N_{\rm{q}}$ is the total number of PAO events, and $m_{ \rm{src},\emph{p} }$ and $m_{ \rm{bkg},\emph{p} }$ are the pixelized maps that were discussed in Appendix~\ref{sec:LikeApp}. As  $N_{\rm{q}}=69$, the function becomes extremely steep in $\Gamma$ and $\emph{R}$, as Figure~\ref{fig:PoissNInt} shows. For such a posterior, $B_{12}$ tends to zero as ${\rm Pr}(\Gamma = 0 |\textbf{\emph{d}}, M_2) \ll 1$.

\subsection{On/Off likelihood}

An additional case that is of interest in this analysis is that of On/Off measurements. In high-energy astrophysics, when a measurement is taken of the number of counts coming from a source of interest, often an auxiliary measurement is made by pointing the detector off-source. These are called the On and Off measurements, respectively. The counts that are detected in the Off measurement are thereby produced solely by the background rate $R$, while the counts in the On measurement are produced by both the background and the source rates $\Gamma$ and $R$. From these two measurements, the source rate can then be estimated (e.g. \citealt{Gregory2010}). 
\par
The likelihood for these kinds of measurements is the product of the Poisson likelihoods of the On and Off measurements:
\begin{equation}
\label{eq:OOLikel}
\rm{Pr(} \emph{N}_{\rm{on}},\emph{N}_{\rm{off}} | \Gamma , \emph{R} \rm{)} = \frac{ ( \emph{R}  \emph{T}_{\rm{off}} )^{\emph{N}_{\rm{off}}} \exp(-\emph{R} \emph{T}_{\rm{off}} )  }{\emph{N}_{\rm{off}}!}  \times \frac{ [(\Gamma + \emph{R})  \emph{T}_{\rm{on}}]^{\emph{N}_{\rm{on}}} \exp[ -(\Gamma + \emph{R})  \emph{T}_{\rm{on}} ] }{\emph{N}_{\rm{on}}!}   ,
\end{equation} 
where $\emph{N}_{\rm{on}}$ and $\emph{N}_{\rm{off}}$ are the numbers of counts on and off source, and $\emph{T}_{\rm{on}}$ and $\emph{T}_{\rm{off}}$ times the detector spends on and off the source. An example of such a likelihood is displayed in Figure~\ref{fig:OOpics}. The On/Off likelihood is very similar to the Poisson product likelihood, as the former can be regarded as a special case of the latter. Thus, the dependence of the Bayes factor on $s$ can be expected to be similar to the dependence for the Poisson product case.
\par 
For the On/Off case, a standard expression for the Bayes factor has been derived (\citealt{Gregory2010}), and can be written as
\[
B_{21}=
\]
\begin{equation}
\label{eq:OOBayes}
\frac{\emph{N}_{\rm{on}}!}{\Gamma_{\rm{max}}\emph{T}_{\rm{on}}\gamma[(\emph{N}_{\rm{on}}+\emph{N}_{\rm{off}}+1],\emph{R}_{\rm{max}}(\emph{T}_{\rm{on}}+\emph{T}_{\rm{off}})]} \sum^{\emph{N}_{\rm{on}}}_{i=0} \frac{ \gamma[(\emph{N}_{\rm{on}}+\emph{N}_{\rm{off}}-i+),\emph{R}_{\rm{max}} (\emph{T}_{\rm{on}}+\emph{T}_{\rm{off}})  ]   }{i!(\emph{N}_{\rm{on}}-i)!} \gamma(i+1,\Gamma_{\rm{max}}\emph{T}_{\rm{on}})\left(1 + \frac{\emph{T}_{\rm{off}}}{\emph{T}_{\rm{on}}} \right)^i  ,
\end{equation} 
where $\gamma(s,x)$ is the lower incomplete gamma function, defined here as
\begin{equation}
\label{eq:OOLikel}
\gamma(s,x) = \int_0^x t^{s-1}e^{-t} \rm{d} \emph{t}  .
\end{equation}
The standard expression reproduces the same dependence that one obtains by calculating the ratio in Equation~\ref{eq:SDDRApp}. Note that this expression is for $B_{21}$ rather than $B_{12}$.

Figure~\ref{fig:BayesF2} shows the dependence of the Bayes factor on $s$ for the three cases. The Bayes factors that are shown in the Figure are the Bayes factors favouring the complex model, $B_{21}=1/B_{12}$. For all three cases, $B_{21}$ falls linearly for $s>1$. For lower values of $s$, the Bayes factor  for the Gaussian case approaches 1, while for the PAO and On/Off cases $B_{12}$ becomes $\gg 1$, as the uniform model is extremely disfavoured. The Bayes factors for the On/Off case behave very similarly to the Poisson product case, as the former can be regarded as a special case of the later.

\begin{figure*}
  \centering $
\arraycolsep=0.01pt\def\arraystretch{-0.5}
\begin{array}{ccc}
\includegraphics[width=85mm]{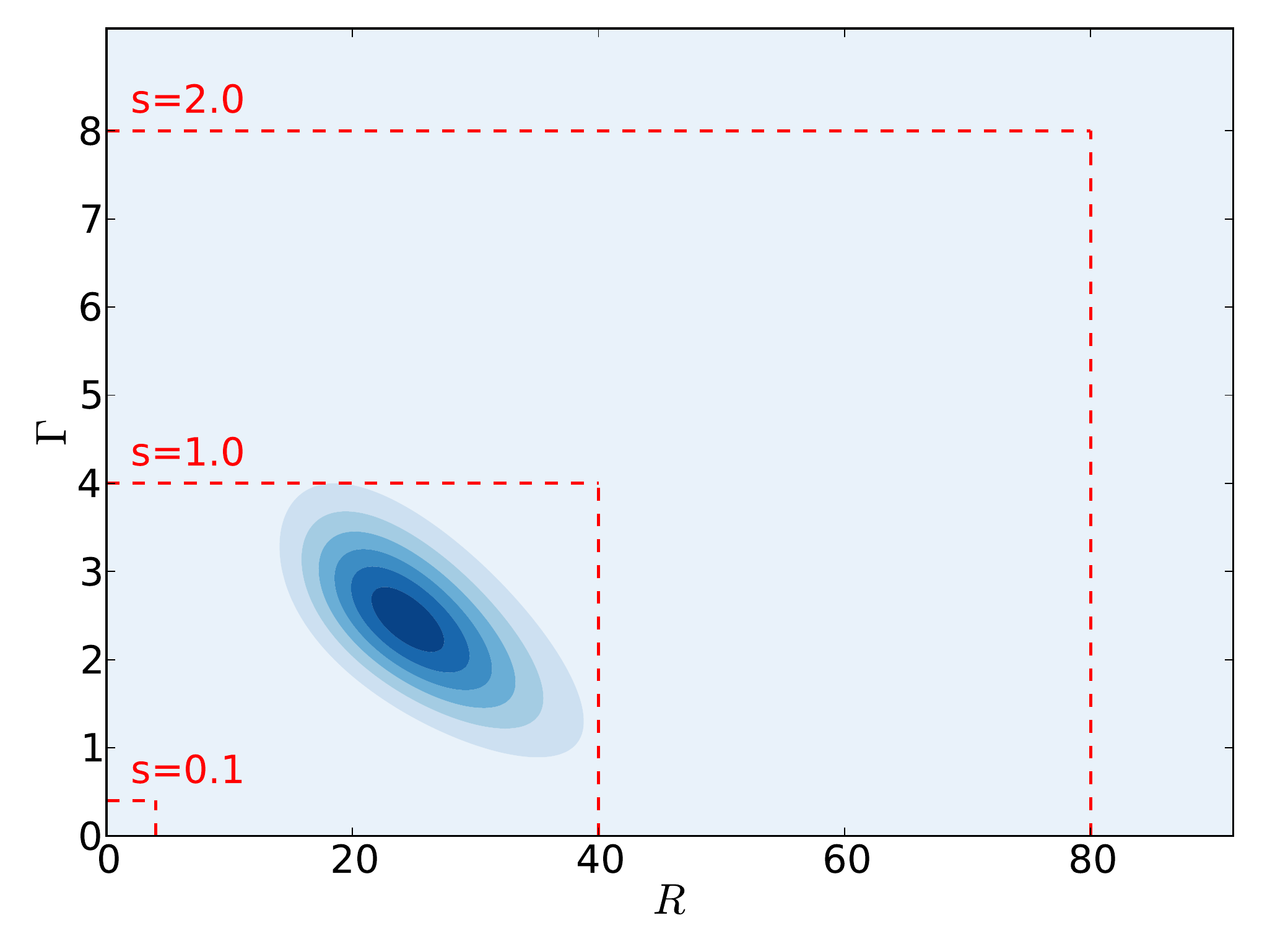}&
\end{array}$
\label{fig:OOpics}
\end{figure*}

\begin{figure}
\begin{center}$
\arraycolsep=0.000000000000000000000pt
\begin{array}{ccc}
\hbox{\includegraphics[width=62mm]{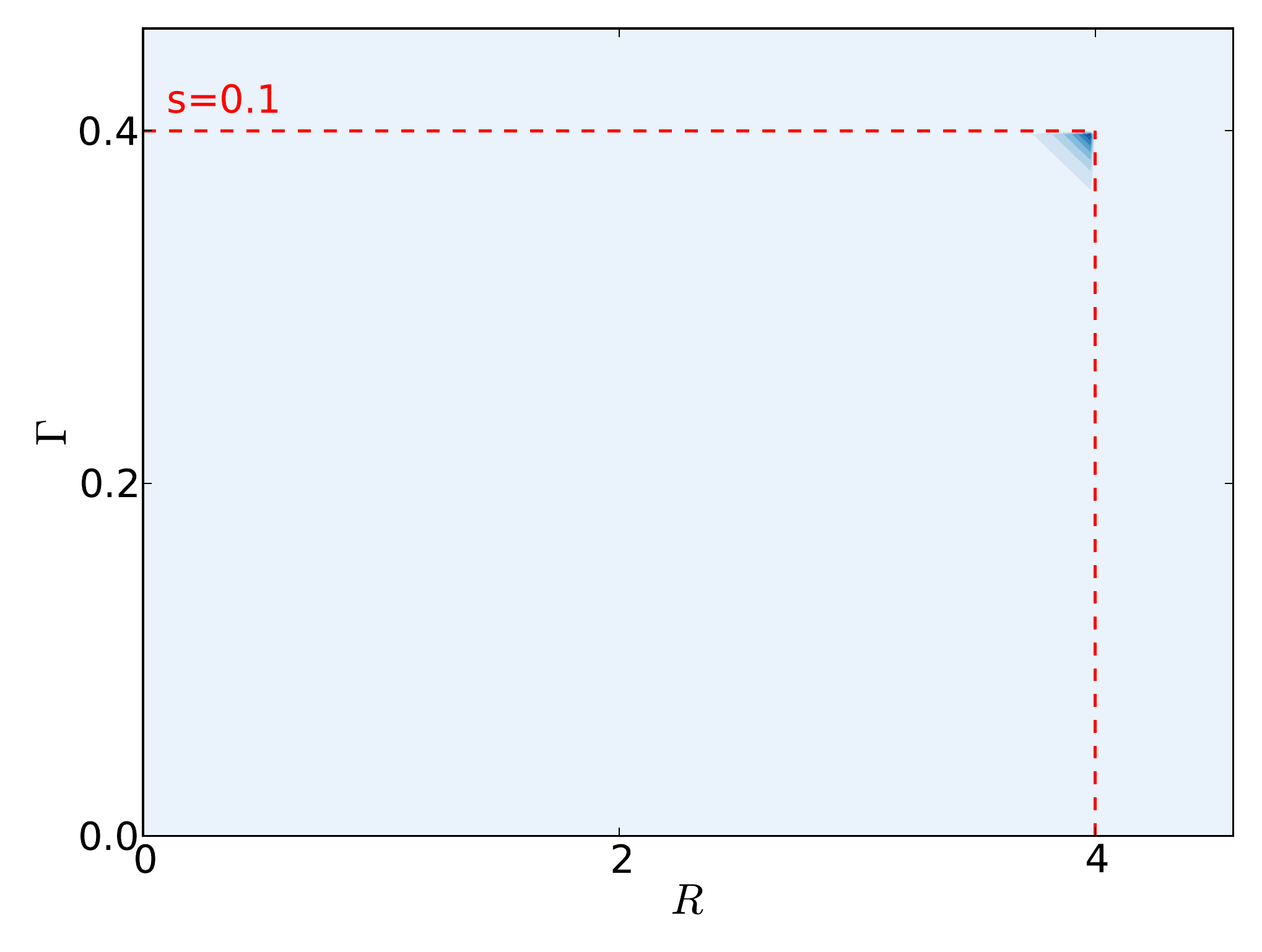}} & \includegraphics[width=62mm]{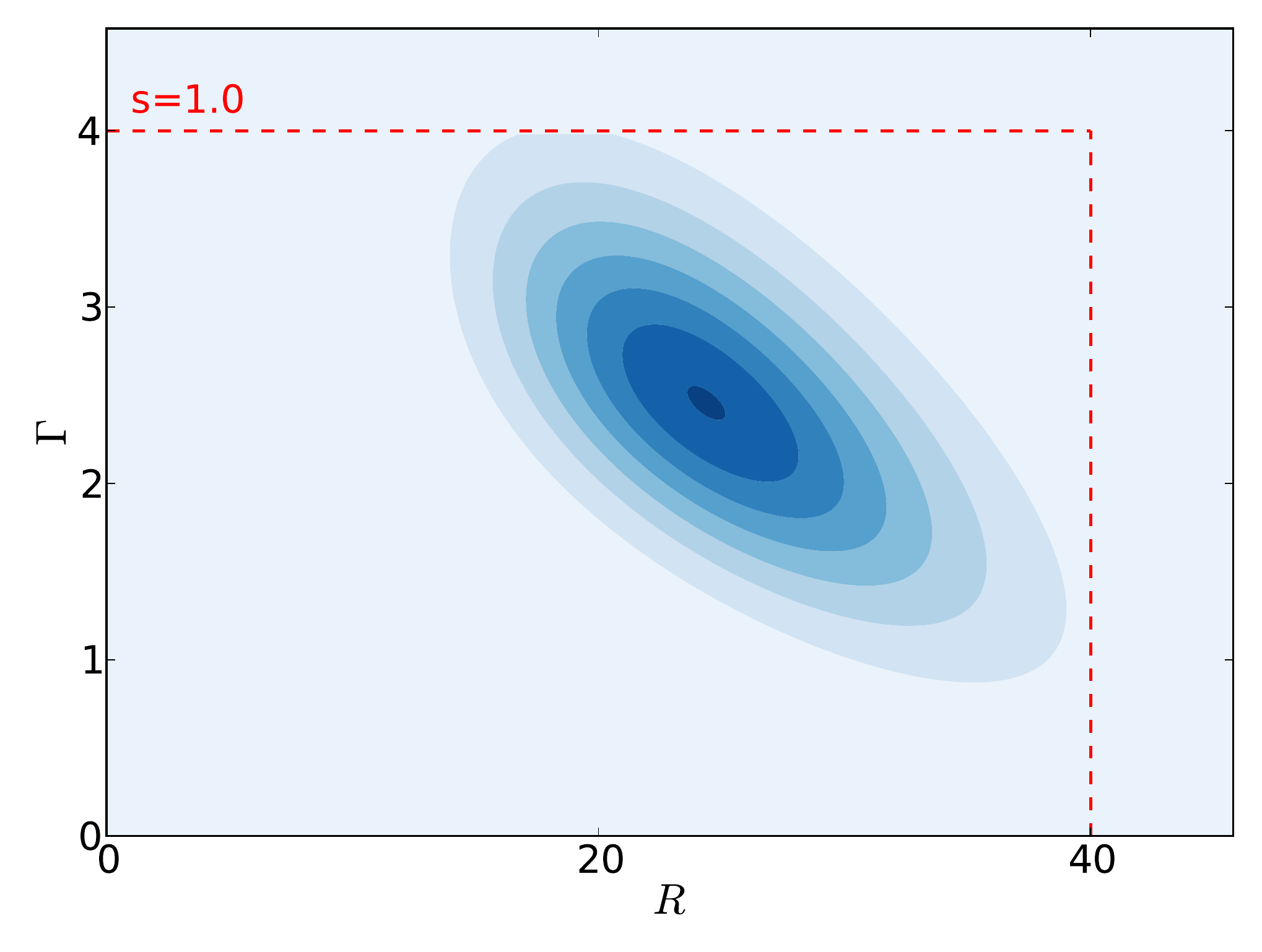} & \includegraphics[width=62mm]{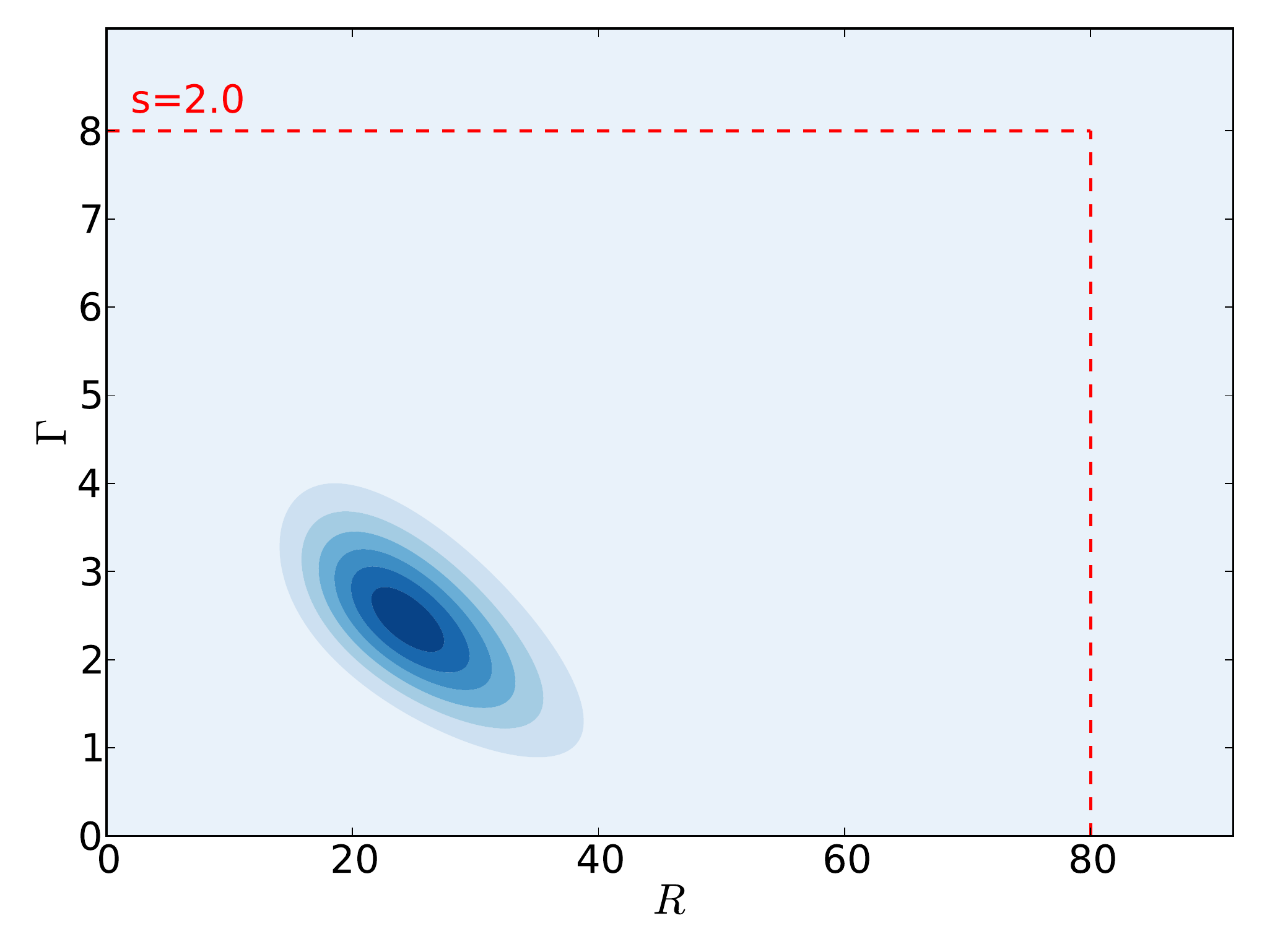} \\
\end{array}$
\end{center}
 \caption{Upper panel: Example On/Off likelihood. The red lines denote prior regions for three different values of the hyperparameter $s$. Lower panel: Posteriors for the same $s$ values are displayed. }
\label{fig:OOpics}
\end{figure}
\par
\begin{figure}
\begin{center}
\hbox{\includegraphics[width=100mm]{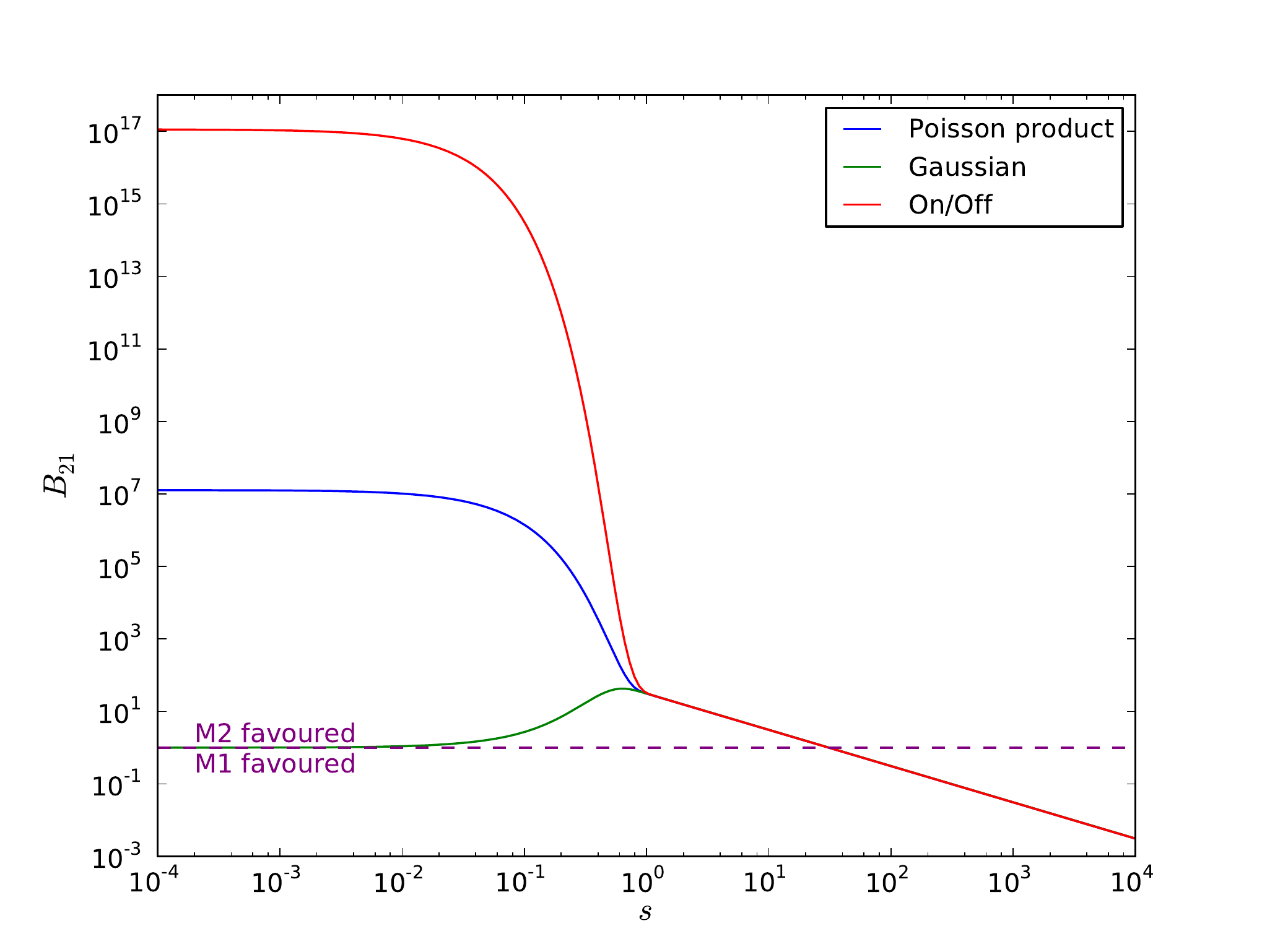}}
\end{center}
\hspace{5cm}
\vspace{-1cm}
\caption{Dependence of the Bayes factor on hyperparameter $s$ for the cases of Gaussian likelihood, the Poisson product likelihood, and the On/Off likelihood. } 
\label{fig:BayesF2}
\end{figure}
\par







 


\bsp
\label{lastpage}
\end{document}